\documentclass[11pt]{article}

\usepackage{bbm}
\usepackage{amsfonts}
\usepackage{amsmath}
\usepackage{amssymb}
\usepackage{amsthm}
\usepackage[mathscr]{eucal}
\usepackage{bbold}
\usepackage{braket}
\usepackage{color}
\usepackage{dsfont}
\usepackage{framed}
\usepackage{jheppub}
\usepackage{mathtools}
\usepackage{physics}
\usepackage[normalem]{ulem}
\usepackage{tensor}
\usepackage{thmtools}
\usepackage{thm-restate}
\usepackage{ulem}
\usepackage{tikz}
\usepackage{soul}

\usetikzlibrary{math} 

\definecolor{gcolor}{RGB}{0,100,0}

\definecolor{rcolor}{RGB}{200,0,0}

\definecolor{bcolor}{RGB}{10,10,255}

\newcommand{\ignore}[1]{}

\makeatletter\def\@fpheader{~}\makeatother
\newcommand{\Eqref}[1]{Eq.~\eqref{#1}}
\newcommand{\secref}[1]{Sec.~(\ref{#1})}
\newcommand{\figref}[1]{Fig.~(\ref{#1})}
 
\usepackage[export]{adjustbox}

\title{Large and Small Corrections to the JLMS Formula from Replica Wormholes}
 
\author[1,2]{Jonah Kudler-Flam}
\emailAdd{jkudlerflam@uchicago.edu}

\author[3]{and Pratik Rath}
\emailAdd{rath@ucsb.edu}

\affiliation[1]{Kadanoff Center for Theoretical Physics, University of Chicago, Chicago, IL~60637, USA}
\affiliation[2]{Department of Physics, Princeton University, Princeton, NJ~08540, USA}
\affiliation[3]{Department of Physics, University of California, Santa Barbara, CA 93106, USA}

\abstract{
The JLMS formula relates the bulk and boundary relative entropies and is fundamental to the holographic dictionary, providing justification for entanglement wedge reconstruction. We revisit the replica trick for relative entropy and find corrections to the JLMS formula in a variety of scenarios, even after accounting for effects of quantum extremality. We analyze the problem in the PSSY model, a model of Jackiw-Teitelboim gravity coupled to end-of-the-world branes. We find non-perturbative (in $G$) corrections that are always present, arising from subdominant replica wormhole gravitational saddles that indicate the approximate error-correcting nature of AdS/CFT. Near entanglement phase transitions, these saddles can get enhanced to large corrections.  We find $O\left(G^{-1/2}\right)$ corrections arising from area fluctuations and $O\left(G^{-1}\right)$ corrections from incompressible bulk quantum states. Lastly, we find our most surprising result, an \textit{infinite} violation of the JLMS formula after the Page time arising from a rank deficiency in the bulk entanglement spectrum. We discuss similar calculations in tensor networks and comment on the implications for bulk reconstruction.}

\begin{document}
 
 
\maketitle

\section{Introduction}\label{sec:intro}

Investigating the AdS/CFT correspondence with a quantum-information perspective has led to tremendous progress in understanding black holes and quantum gravity. The starting point for this fruitful pursuit is the Ryu-Takayanagi (RT) formula with FLM quantum corrections \cite{2006PhRvL..96r1602R,2006JHEP...08..045R,2007JHEP...07..062H,2013JHEP...11..074F}:
\begin{equation}\label{eq:FLM}
	S(\rho_R) = \frac{\mathcal{A}(\gamma_R)}{4G} + S_{\text{bulk}}(\rho_r),
\end{equation}
where $R$ is a boundary subregion and $\gamma_R$ is the minimal-area surface anchored to $\partial R$. $\rho_r$ is the reduced density matrix on a bulk subregion $r$ such that $\partial r = \gamma_R \cup R$, known as the \textit{entanglement wedge} of $R$. \Eqref{eq:FLM} consequently led to a formula relating the bulk and boundary relative entropy, commonly known as the JLMS formula \cite{2016JHEP...06..004J}:
\begin{equation} \label{eq:JLMS}
	D(\rho_R||\sigma_R) = D_{\text{bulk}}(\rho_r||\sigma_r),
\end{equation}
where $D(\rho||\sigma)$, the relative entropy, is a measure of distinguishability of quantum states defined as
\begin{align}
    D(\rho||\sigma) = \Tr\left[\rho \log\left[\rho\right] \right] -\Tr\left[\rho \log\left[\sigma\right] \right].
\end{align}

While the arguments leading up to \Eqref{eq:FLM} and \Eqref{eq:JLMS} used the Euclidean path integral \cite{2013JHEP...11..074F,2013JHEP...08..090L}, it was subsequently given a Lorentzian, Hilbert space interpretation by understanding the holographic dictionary as a quantum error-correcting code (QEC) \cite{2015JHEP...04..163A,2016PhRvL.117b1601D,2017CMaPh.354..865H}. An exact QEC is an isometric map from the bulk effective field theory (EFT) Hilbert space to the boundary Hilbert space such that it spans a ``code" subspace whose information is encoded redundantly on the boundary. It was demonstrated that both \Eqref{eq:FLM} and \Eqref{eq:JLMS} are straightforward consequences of the existence of such an exact QEC \cite{2017CMaPh.354..865H}. In fact, the JLMS formula implies the existence of a bulk reconstruction map, the \textit{Petz map}, that can be used to write down bulk operators in terms of their boundary representation \cite{ohya2004quantum,2017arXiv170405839C,2020arXiv200804810F}.

While the above results are quite illuminating, they are by no means the complete story. An emerging theme over the past few years has been that the holographic map is only an approximate QEC.\footnote{See Ref.~\cite{2021arXiv211014669K} for a recent review.} Non-perturbative corrections from the gravitational path integral result in errors in the encoding map, often leading to qualitatively novel features that do not arise for exact QECs \cite{2019JHEP...12..007H}.

In particular, arguments using the path integral have updated the RT formula with quantum corrections to the quantum extremal surface (QES) formula \cite{2015JHEP...01..073E}. Interestingly, the QES formula can also be given a Lorentzian interpretation by understanding the holographic map as an approximate QEC \cite{2021arXiv210914618A}.\footnote{We will work with stationary spacetimes and thus, we will only be interested in the minimality in the spatial direction. Extremality in the timelike direction hasn't so far been explained satisfactorily from QEC.} The QES formula has provided remarkable insights into the black hole information problem, where it has explained the Page curve of black hole evaporation \cite{2019JHEP...12..063A,2020JHEP...09..002P,2020JHEP...03..149A}. 

Once one considers the QES formula, the JLMS formula is updated to a version we refer to as the quantum extremal JLMS (qJLMS) formula\footnote{The first term in this formula is confusing when the geometries describing $
\rho$ and $\sigma$ are macroscopically different because it is unclear how the $\gamma_{\sigma}$ surface should be defined in the $\rho$ geometry. We circumvent this ambiguity by always taking $\rho$ and $\sigma$ to have the same bulk geometries.}
\cite{2018JHEP...01..081D}
\begin{align}\label{eq:QEJLMS}
    D(\rho_R ||\sigma_R) &= \left\langle  \frac{\hat{\mathcal{A}}^{\gamma_{\sigma}}}{4G}-\log \sigma_r\right\rangle_{\rho} - \left\langle  \frac{\hat{\mathcal{A}}^{\gamma_{\rho}}}{4G}-\log \rho_r\right\rangle_{\rho},
\end{align}
where $\gamma_{\sigma}$ ($\gamma_{\rho}$) are the codimension-2 surfaces extremizing the first (second) terms in the above equation. For the second term, this is the QES while for the first term, it is the modular extremal surface defined in Ref.~\cite{2018JHEP...01..081D}. $\hat{\mathcal{A}}$ is the \textit{area operator}, which is a linear operator defined on the gravitational phase space that evaluates the area of the given surface. Further, the RT formula relates it to the entropy as long as one chooses a sufficiently small code subspace \cite{2017JHEP...02..074A}. 

The qJLMS formula arises from a Euclidean path integral analysis under the assumption of replica symmetry. However, in recent times, much has been learned by performing a more careful analysis of the gravitational path integral including replica symmetry breaking contributions \cite{2019arXiv191111977P,2020JHEP...11..007D,2020JHEP...12..084M,2021JHEP...04..062A,2021PRXQ....2c0347S,2021JHEP...06..024D,2021PhRvL.126q1603K,2021arXiv210505892W,2021PRXQ....2d0340K,2022JHEP...02..076K,2021arXiv211011947D,2021arXiv211200020V,2021arXiv211002959V,2021arXiv211209122A,2022arXiv220111730A}. We are therefore motivated to revisit the calculation of the relative entropy using the path integral and analyze corrections to the qJLMS formula. As expected, we find perfect agreement with qJLMS in all cases when evaluating the contributions from only the replica symmetric saddles. However, including all the remaining contributions to the path integral, we find small non-perturbative corrections as well as large corrections in certain situations.

In order to do so, we will analyze this problem in the ``PSSY model,'' a toy model of black hole evaporation where the gravitational path integral is tractable \cite{2019arXiv191111977P}. The model is that of Jackiw-Teitelboim (JT) gravity decorated with end-of-the-world (ETW) branes that carry non-dynamical flavor indices. Owing to its simplicity, this model has been tremendously useful for understanding non-perturbative effects that become important during ``phase transitions'' in various quantities.

\begin{figure}
    \centering \includegraphics[width=0.8\textwidth]{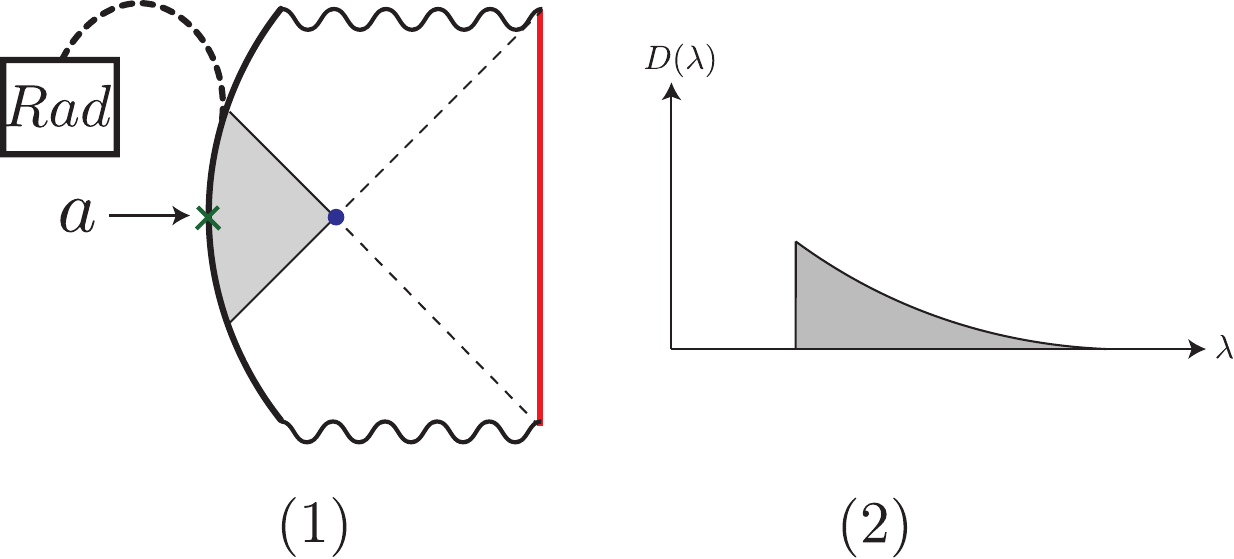}
    \caption{1) The Lorentzian description of the states we consider in the PSSY Model, a JT gravity black hole with an ETW brane. The ETW brane carries $k$ flavor indices that are entangled (dashed lines) with the radiation system labelled $Rad$. There are additional indices (labelled $a$) that span the code subspace of bulk states we consider. The extremal surface is denoted in purple and the island that dominates after the Page time is coloured gray. 2) The spectrum of the boundary density matrix $\rho_{Rad}$ post Page time is a cutoff thermal spectrum without any zero eigenvalues when the black hole is in the canonical ensemble. Thus, the boundary relative entropy is finite while the bulk relative entropy is infinite for distinct pure states.}
    \label{fig:west_coast_intro}
\end{figure}

In the PSSY model, we first define a code subspace of states by considering additional flavor indices on the ETW brane as depicted in \figref{fig:west_coast_intro}. We then calculate the boundary relative entropy between various choices of bulk states by employing a replica trick calculation. Doing so, we find the following classes of corrections to the qJLMS formula:
\begin{itemize}
	\item $O(e^{-1/G})$: Non-perturbative corrections arise from the existence of subdominant saddles. Such configurations exist in the presence of non-minimal QESs.
	\item $O(G^{-1/2})$: Such corrections, which have appeared previously in the context of entanglement entropy phase transitions \cite{2017PhRvL.119v0603V,2019PhRvE.100b2131M,2019arXiv191111977P,2020JHEP...11..007D,2020JHEP...12..084M}, also arise for the relative entropy in a similar manner. The key mechanism leading to these corrections is $O(G^{-1/2})$ energy fluctuations in the canonical ensemble.
	\item $O(G^{-1})$: Leading order corrections to the QES formula are now well understood to arise when considering bulk states that are \textit{incompressible} \cite{2021JHEP...04..062A, 2021arXiv210505892W}. The failure of the QES formula, which is the central ingredient entering \Eqref{eq:QEJLMS}, naturally leads to large corrections of $O(G^{-1})$ to the qJLMS formula as well.  
	\item \textit{Infinite}: The most novel, surprising violations of the qJLMS formula arise when considering pure states in the bulk. Despite the bulk relative entropy being universally \textit{infinite} for distinct pure states, the boundary quantum system that encodes the Hawking radiation has full rank, leading to a \textit{finite} boundary relative entropy. The existence of energy fluctuations plays an important role for such violations of the qJLMS formula. In particular, exponentially small tails in the wavefunction that arise from macroscopically large fluctuations result in an avoidance of \textit{rank deficiency} for the boundary quantum system.
\end{itemize}
The above classes of corrections are rather generic and we expect similar features to arise in more general models of holography from analogous mechanisms.\footnote{It was argued in Ref.~\cite{2018JHEP...01..081D} that qJLMS holds to all orders in $G$ under the assumption of replica symmetry. In principle, replica symmetry breaking contributions could lead to perturbative corrections at all orders in $G$ as well. They do not show up in this simple model which is one-loop exact \cite{2017JHEP...10..008S}, and summing up non-perturbatively suppressed saddles is well justified.} 

We would like to emphasize that although the first three mechanisms of corrections have been seen previously in other related settings, the infinite violations we find are qualitatively new. They essentially arise from the fact that the relative entropy changes discontinuously under small changes to the state, unlike stable quantities like the von Neumann entropy and reflected entropy \cite{cmp/1103859037,2007JPhA...40.8127A, 2020JHEP...04..208A}. As shown in \figref{fig:west_coast_intro}, the spectrum of the boundary density matrix is cutoff so that it doesn't have any zero eigenvalues. While this doesn't significantly affect the state as measured by trace distance, it ensures that the boundary relative entropy is finite. On the other hand, the bulk relative entropy is infinite for distinct pure states, thus leading to an infinite violation of the qJLMS formula.\footnote{We note that the violation can be made arbitrarily large without being infinite if all zero eigenvalues are replaced by arbitrarily small but positive eigenvalues.} 

This discontinuous behaviour also extends to the modular Hamiltonian, which is the logarithm of the density matrix. Despite the fact that the bulk and boundary density matrices are related by an approximate isometry in our examples, we find that the operator version of the qJLMS formula that relates the bulk and boundary modular Hamiltonians is also violated by a large amount. 

Having found these corrections, we discuss the implications of corrections to the qJLMS formula. An important role that the JLMS formula plays in holography is that of justifying bulk reconstruction and subregion duality. In particular, the JLMS formula being satisfied to a high degree implies the existence of a reconstruction map \cite{2015arXiv150907127J, 2017arXiv170405839C,2020JHEP...01..168C}, i.e.,
\begin{align}
    D(\rho_r || \sigma_r ) - D(\rho_R||\sigma_R)\geq -2\log\left[ F(\rho_r, \mathcal{R}_{\sigma_r,\mathcal{N}}\circ \mathcal{N}[\rho_r])\right],
    \label{eq:junge}
\end{align}
where $\mathcal{N}(\rho_r)$ is the completely-positive, trace-preserving map that maps the bulk density matrix $\rho_r$ to a boundary density matrix $\rho_R$, and $\mathcal{R}$ is a recovery map, independent of $\rho_r$, that attempts to recover the information in bulk subregion $r$ on boundary subregion $R$. If the JLMS formula is satisfied up to small errors, there exists a recovery map that works with good fidelity \cite{2017arXiv170405839C,2020JHEP...01..168C}.\footnote{These results have also been generalized to the case of Type III${}_1$ von Neumann algebras where the Hilbert space does not factorize into subregions \cite{2020arXiv201005513F,2021arXiv211212789G,2020arXiv200608002F}.}

However, the degree to which the JLMS formula holds only provides a lower bound on the fidelity of recovery. Thus, not all corrections to the JLMS formula lead to a failure of the bulk reconstruction map. For example, while the $O(G^{-1})$ corrections are expected to lead to a failure of bulk reconstruction since the entanglement wedge is not well defined \cite{2021JHEP...04..062A}, the infinite violations we find arise in the vanilla situation where the entanglement wedge is sharply defined and one should expect bulk reconstruction to work with good fidelity. We show that this is indeed the case by identifying a high-fidelity reconstruction map. The violations we find essentially arise from the sensitivity of the relative entropy to small changes in the state, and is not always the optimal diagnostic of the fidelity of reconstruction.

We now provide a brief overview of the paper. In \secref{sub:pssy}, we review the PSSY model and define a code subspace of bulk states for our analysis by using a path integral description. In \secref{sub:replica}, we set up the replica trick calculation of relative entropy in this model. In \secref{sub:qjlms}, we show how replica symmetry leads to the qJLMS formula in the PSSY model. We then compute the relative entropy in various cases by including replica-symmetry breaking contributions in \secref{sec_violations} to find corrections to the qJLMS formula. The corrections are classified as discussed above, and we illustrate them with an example of each type. Finally, we conclude with some implications of our work and directions for future research in \secref{sec:discuss}. We include extensions of our work, such as a resolvent trick for relative entropy and the general solution to relative entropy in random tensor networks, in the Appendices.

\section{Relative Entropy in the PSSY Model}
\label{PSSY_sec}

In this section, we set up the background material for our results in \secref{sec_violations}. In \secref{sub:pssy}, we first review the PSSY model of black hole evaporation which is the setting where we perform our analysis. In particular, we define a code subspace of bulk states in the model. We then setup the replica trick calculation for relative entropy between two such states in \secref{sub:replica}. Finally, we describe how replica symmetry leads to the qJLMS formula in the PSSY model in \secref{sub:qjlms}.

\subsection{PSSY Model}
\label{sub:pssy}

The PSSY model was introduced as a simple model of black hole evaporation where explicit calculations of the entropy of Hawking radiation could be evaluated \cite{2019arXiv191111977P}. The model consists of JT gravity decorated with ETW branes that carry a large number of flavor indices (we will take this integer to be $d_{code}\times k$). The total gravitational action is given by
\begin{align}
    I = -\frac{S_0}{2\pi}\left[\frac{1}{2}\int_{\mathcal{M}}\sqrt{g}R+ \int_{\partial \mathcal{M}}\sqrt{h}K \right]-\left[\frac{1}{2}\int_{\mathcal{M}}\sqrt{g}\phi(R+2)+\int_{\partial \mathcal{M}}\sqrt{h}\phi K \right] +\mu \int_{brane}ds,
\end{align}
where $S_0$ is the (large) extremal entropy, $g$ ($h$) is the bulk (asymptotic boundary) metric with intrinsic (extrinsic) curvature $R$ ($K$). $\phi$ is the dilaton, and $\mu$ is the mass of the ETW brane.

In order to consider a code subspace of bulk states, we divide the flavor indices in two: an index $i \in \{1,\dots,k\}$ and an index $a \in \{1, \dots, d_{code}\}$. A basis for the $d_{code}$ dimensional code subspace is defined by entangling the $k$ flavor indices with an auxiliary ``radiation'' system $Rad$:
\begin{align}\label{eq:code}
    \ket{\Psi^{(a)}} = \frac{1}{\sqrt{k}}\sum_{i = 1}^k \ket{i}_{Rad}\ket{\psi_{ia}}_B, \quad a \in \{1, \dots, d_{code}\},
\end{align}
where $\ket{\psi_{ia}}_B$ is the state of the black hole with frozen flavor index~$i$ and fluctuating index~$a$. $\ket{\psi_{ia}}_B$ may be prepared using the Euclidean gravitational path integral. It is useful to consider a diagrammatic representation of these states. An overlap between two states defines the following boundary condition
\begin{equation}\label{eq:overlap}
	\langle \psi_{jb}|\psi_{ia}\rangle = \ \includegraphics[scale = .7,valign = c]{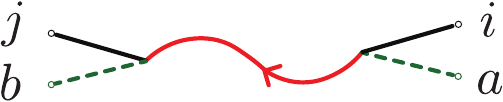},
\end{equation}
where the red line represents the asymptotic boundary of the spacetime, the black lines represent the frozen flavor indices and the dashed green lines represent the fluctuating flavor indices defining the code subspace. To evaluate the overlap, we must sum over all bulk configurations consistent with these boundary conditions, e.g.
\begin{equation}\label{eq:euc_geom}
	\langle \psi_{jb}|\psi_{ia}\rangle = \ \includegraphics[scale = .7,valign = c]{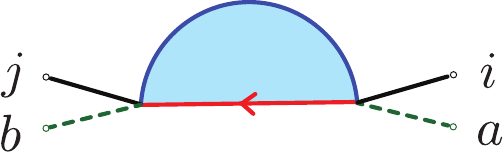},
\end{equation}
The extension of the brane into the bulk (denoted by a blue line) gives it a definite flavor, causing the diagram to be proportional to $\delta_{ij} \delta_{a b}$. 

\begin{figure}
    \centering \includegraphics[width=0.9\textwidth]{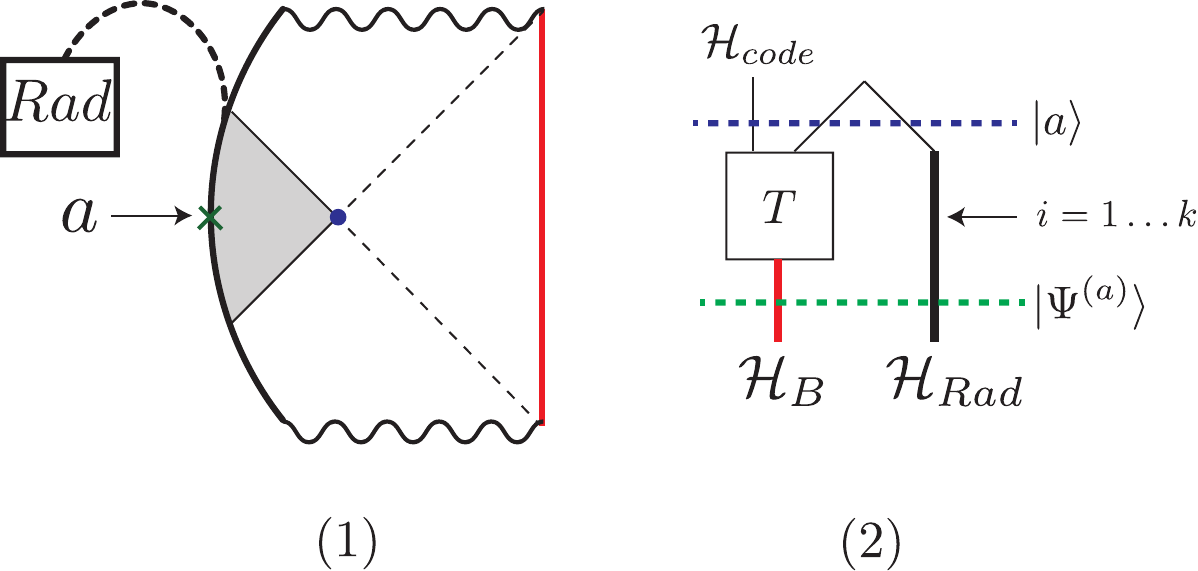}
    \caption{1) The Lorentzian description of a state $\ket{\Psi^{(a)}}$ in the code subspace we consider in the PSSY Model, a JT gravity black hole with an ETW brane. The ETW brane carries frozen flavor indices that are entangled (dashed black line) with radiation system $Rad$. They also carry fluctuating flavor indices, denoted $a$, that can be arbitrarily specified. The extremal surface is denoted in purple and the island that dominates for sufficiently large $k$ is coloured gray. 2) A tensor network description of the states in the code subspace. The boundary state $\ket{\Psi^{(a)}}$ is defined in terms of cut of a tensor network (dashed green line). The tensor $T$ represents the encoding map defined by \Eqref{eq:dual}. The semiclassical state $\ket{a}$ is defined by a different cut of the tensor network (dashed blue line). The network represents an approximate isometry from $\mathcal{H}_{code}$ to $\mathcal{H}_{B}\otimes \mathcal{H}_{Rad}$.}
    \label{fig:wc}
\end{figure}

One can analytically continue the Euclidean geometry in \Eqref{eq:euc_geom} about the moment of time symmetry to obtain a Lorentzian solution of a black hole with an ETW brane behind the horizon with flavor indices $i$ and $a$. In addition, the states in the code subspace defined in \Eqref{eq:code} involve the frozen indices being entangled with a radiation system $Rad$. Combining these, the Lorentzian description of the states in our code subspace is given by \figref{fig:wc}. 

Varying the parameter $k$ allows us to understand the Page transition. $k<e^{S_{BH}} d_{code}$, where $S_{BH}$ is the black hole entropy, corresponds to early stages of evaporation whereas $k>e^{S_{BH}} d_{code}$ represents the period of evaporation beyond the Page time. It is by now well understood that the entanglement wedge of $Rad$ includes an island in the interior of the black hole post-Page time \cite{2019JHEP...12..063A,2020JHEP...09..002P,2020JHEP...03..149A}. Thus, by entanglement wedge reconstruction, $Rad$ has access to the quantum information of the fluctuating flavors which span the code subspace.

For our calculations, we will consider two distinct asymptotic boundary conditions. These are fixed energy, $E := s^2/2$,  and fixed renormalized length, $\beta$, corresponding to the microcanonical and canonical ensemble respectively. Partition functions with $n$ asymptotic boundaries with a connected bulk with these boundary conditions have been evaluated in Ref.~\cite{2019arXiv191111977P}. In the microcanonical ensemble, the partition function has a simple exponential dependence on $n$:
\begin{align}
    Z_{n}(E) = e^{\textbf{S}}y(E)^n , \quad y(s) = e^{-\frac{\beta s^2}{2}}2^{1-2\mu}|\Gamma(\mu -\frac{1}{2}+is)|^2,
\end{align}
where $\textbf{S}$ is the microcanonical entropy at energy $E$. In the canonical ensemble, there is a representation of the partition function as an integral over energies:
\begin{align}
    Z_n(\beta) = e^{S_0}\int_0^{\infty}ds \varrho(s) y(s)^n, \quad \varrho(s) = \frac{s}{2\pi^2}\sinh(2\pi s).
\end{align}

Having described the model of gravity, we would like to mention that the holographic dual of JT gravity is well known to be an ensemble average over random Hamiltonians \cite{2019arXiv190311115S}. Similarly, in Ref.~\cite{2019arXiv191111977P}, the PSSY model was shown to have a holographic dual consisting of an ensemble average, now including additional Gaussian random variables for the flavor indices. In more detail, the states $\ket{\psi_{ia}(\beta)}$ in the canonical ensemble are given by
\begin{equation}\label{eq:dual}
	\ket{\psi_{ia}(\beta)} = \sum_{j} 2^{1/2-\mu} \Gamma(\mu-\frac{1}{2} +i \sqrt{2 E_j})e^{-\beta E_j/2} C_{i a,j} \ket{E_j}, 
\end{equation}
where $C_{ia,j}$ are independent complex Gaussian random variables and $\ket{E_j}$ are the eigenstates of a given Hamiltonian. Computations in JT gravity using the path integral then correspond to calculations that average over the Hamiltonian and flavor random variables with the appropriate measure. \Eqref{eq:dual} provides us with a tensor network representation of \Eqref{eq:code}, the states in our code subspace. This serves as a useful visual tool to understand the physical setting. In particular, it helps distinguish between the semiclassical bulk state and the fine grained state as depicted in \figref{fig:wc}.

\subsection{Replica Trick for Relative Entropy}
\label{sub:replica}

Due to the logarithms in its definition, the relative entropy is difficult to evaluate directly. Instead, we will use a replica trick, evaluating certain integer moments of the density matrices and analytically continuing the replica number, analogous to the replica trick for von Neumann entropy \cite{2016PhRvL.117d1601L}
\begin{align}\label{eq:replica}
    D(\rho_{Rad} ||\sigma_{Rad}) = \lim_{n\rightarrow 1}\frac{1}{1-n} \left(\log\left[{\Tr\left[\rho_{Rad}  \sigma_{Rad}^{n-1}\right]} \right]-\log\left[{\Tr\left[\rho_{Rad}^n\right]} \right]\right),
\end{align}
where we refer to the first term as the \textit{relative term} and the second one as the \textit{entropy term}. 

First, we can introduce a diagrammatic representation for the density matrices $\rho_{Rad}$, $\sigma_{Rad}$ based on \Eqref{eq:code} and \Eqref{eq:overlap}. We have
\begin{align}
	\left(\rho_{Rad}\right)_{ij} &= \frac{1}{k} \sum_{a,b=1}^{d_{code}} \langle \psi_{jb}|\psi_{ia}\rangle_B\, \left(\ket{i}\bra{j}\right)_R \rho_{ab}\\
	&=\ \includegraphics[scale = .5,valign = t]{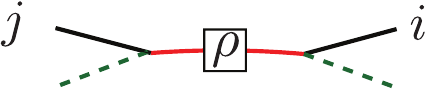},
\end{align}
where we have introduced the box containing $\rho$ and removed the labels for the fluctuating indices as a shorthand for $\sum_{a,b} \rho_{ab}$. Thus, the terms in \Eqref{eq:replica} can be diagrammatically represented by the following boundary conditions (e.g. for $n=4$):\footnote{We thank Vladimir Narovlanksy for pointing out an error in an earlier version of this draft.} 
\begin{align}
 &\Tr\left[\rho_{Rad}  \sigma_{Rad}^{n-1}\right] = \ \includegraphics[scale = .5,valign = c]{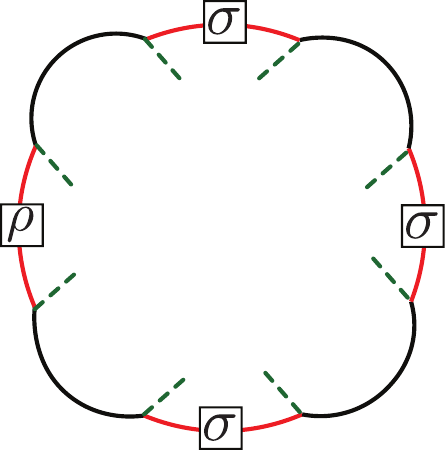}, &\Tr\left[\rho_{Rad}^{n}\right] = \ \includegraphics[scale = .5,valign = c]{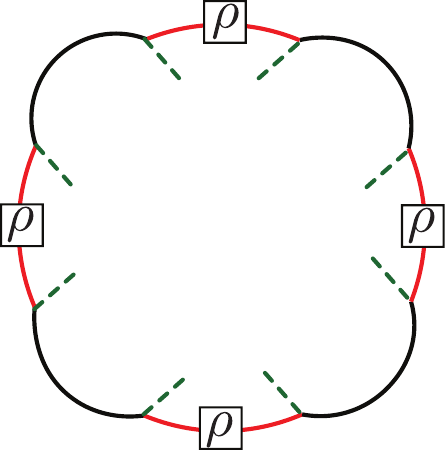}.
\end{align}

Given the boundary conditions, we are then instructed to sum over all bulk configurations consistent with them. The configurations can be classified by their topology. Since configurations with handles in the bulk are suppressed by factors of $e^{-2S_0}$ due to the topological term in the JT action, we will drop these contributions from all calculations. Thus, the topologies are completely classified by the $n!$ different ways to connect the $n$ boundaries with genus-0 geometries, each of which can be associated to an element of the permutation group $S_n$. The topology also dictates the way in which the frozen and fluctuating indices are contracted.

\begin{figure}
    \centering \includegraphics[width=0.6\textwidth]{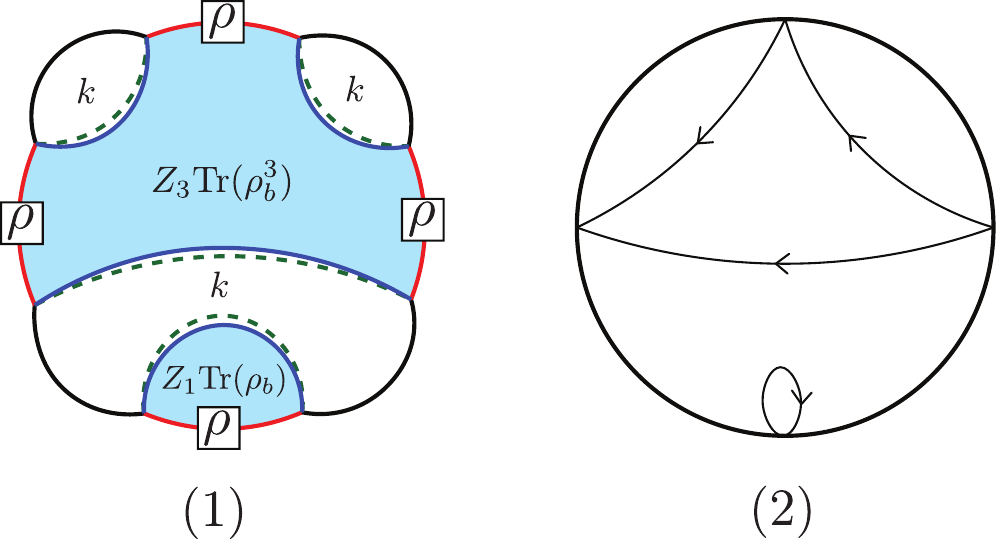}
    \caption{1) The contribution from one of the topologies for the entropy term has been decomposed into the contribution from each section of the planar diagram. Each frozen index loop contributes $k$ while the blue, shaded regions representing a connected disk joining $n$ boundaries contribute $Z_n$ and $\Tr(\rho_{code}^n)$ from the bulk action and the fluctuating index contraction respectively. 2) The same geometry represented as a permutation in the symmetric group $S_4$.}
    \label{fig:eg}
\end{figure}

Represented in terms of permutation group elements, one finds for the entropy term:
\begin{align}
    \Tr\left[\rho_{Rad}^n \right] = \frac{1}{\left(kZ_1\right)^n}\sum_{\tau \in S_n} k^{C\left(\eta^{-1}\circ \tau\right)}\prod_{ i = 1}^{C(\tau)} Z_{n_i} \Tr\left[\rho_{code}^{n_i} \right],
    \label{entropy_term}
\end{align}
where $n_i$ is the length of the $i^{th}$ cycle in permutation $\tau$ and $Z_{p}$ is the JT gravity path integral with $p$ boundaries. $C(g)$ counts the number of cycles, including trivial ones, in the permutation $g$ and $\eta$ is the cyclic permutation. An example of the contribution from a given topology is illustrated in \figref{fig:eg}. Similarly, for the relative term we have:
\begin{align}
    \Tr\left[\rho_{Rad} \sigma_{Rad}^{n-1} \right] = \frac{1}{\left(kZ_1\right)^n}\sum_{\tau \in S_n} k^{C\left(\eta^{-1}\circ \tau\right)}Z_{n_1}\Tr\left[\rho_{code} \sigma_{code}^{n_1 - 1} \right]\prod_{ i = 2}^{C(\tau)} Z_{n_i}\Tr\left[\sigma_{code}^{n_i} \right].
\end{align}

\subsection{The qJLMS Formula}
\label{sub:qjlms}

For each permutation group element, one in fact finds a saddle-point bulk configuration that solves the equations of motion. Among these, there are two special configurations that are \textit{replica symmetric} corresponding to the identity and cyclic permutations. Working in the saddle point approximation under the assumption of replica symmetry, one finds (e.g. for the entropy term for $n=4$) 
\begin{align}
\Tr\left[\rho_{Rad}^n \right] = \text{max}\left[\ \includegraphics[scale = .5,valign = c]{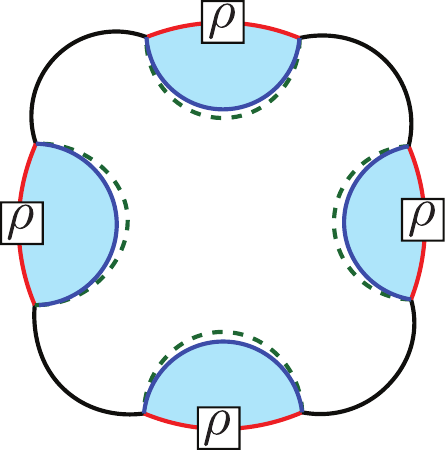}\quad,\quad \includegraphics[scale = .5,valign = c]{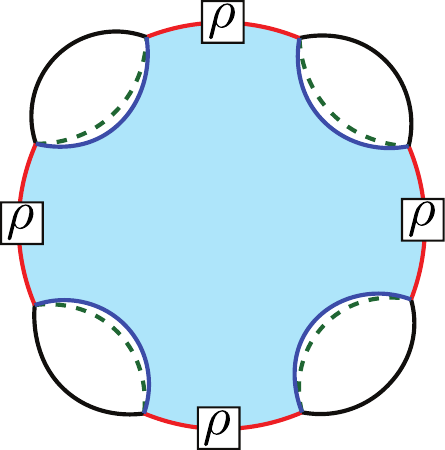}\right],
\end{align}

When the identity element (the fully disconnected geometry) is dominant, the entropy term is identical to the relative term, i.e.,
\begin{align}
    \Tr\left[\rho_{Rad}^n \right] = \Tr\left[\rho_{Rad} \sigma_{Rad}^{n-1} \right] = k^{1-n},
\end{align}
so we manifestly find zero relative entropy. This is the situation for small $k$ or equivalently before the Page time. 

In contrast, when the cyclic permutation (the fully connected term) is dominant,
\begin{align}
    \Tr\left[\rho_{Rad}^n \right] = \frac{Z_n}{Z_1^n} \Tr\left[\rho_{code}^{n} \right], \quad  \Tr\left[\rho_{Rad} \sigma_{Rad}^{n-1} \right] = \frac{Z_n}{Z_1^n} \Tr\left[\rho_{code} \sigma_{code}^{n - 1} \right].
\end{align}
Thus, we find that the relative entropy is in fact equal to bulk relative entropy
\begin{align}
    D(\rho_{Rad} ||\sigma_{Rad}) =  D(\rho_{code} ||\sigma_{code}).
\end{align}

Moreover, the transition in dominance from the identity to cyclic permutations occurs at different locations for the entropy and relative terms determined by the quantum extremality condition as shown in Ref.~\cite{2018JHEP...01..081D}. Therefore, as expected, replica symmetry leads to the qJLMS formula. Similar formulas can be found for the relative entropy at late times in many-body, non-gravitational systems using the so-called equilibrium approximation \cite{2021arXiv211200020V}.

More generally, due to the $\mathbb{Z}_n$ symmetry in the bulk, one may take a quotient by this group to arrive at a geometry with a conical defect. After taking the quotient, there is a natural analytic continuation to real values of $n$, a feature leveraged in the derivations of Refs.~\cite{2013JHEP...08..090L,2013JHEP...11..074F,2018JHEP...01..081D}. This assumption leads precisely to the qJLMS formula as we have confirmed above in the PSSY model. 

\section{Corrections to the qJLMS Formula}
\label{sec_violations}

We have shown that replica symmetry leads to the qJLMS formula in the PSSY model. We will now find corrections to the qJLMS formula by including replica symmetry breaking contributions to the gravitational path integral. We perform computations of relative entropy for various examples in the PSSY model, working through small corrections to large violations of the qJLMS formula.

A natural choice for our examples is to take $\sigma_{code}$ to be the maximally mixed state in the code subspace,
\begin{align}
    \sigma_{code} = \frac{\mathbbm{1}_{code}}{d_{code}}.
\end{align}
The relative term then simplifies to 
\begin{align}
    \Tr\left[\rho_{Rad} \sigma_{Rad}^{n-1} \right] = \frac{1}{\left(kd_{code}Z_1\right)^n}\sum_{\tau \in S_n} k^{C\left(\eta^{-1}\circ \tau\right)}\prod_{ i = 1}^{C(\tau)} d_{code}Z_{n_i}.
\end{align}
This is identical to the entropy term, \Eqref{entropy_term}, for a pure bulk state $\rho_{code}$ and renormalized partition function, $Z_{p}\rightarrow d_{code}Z_p$. Importantly, it is independent of the state $\rho_{code}$. Therefore, the relative entropy may be expressed as the difference between two von Neumann entropies
\begin{align}
    D(\rho_{Rad} || \sigma_{Rad}) = S_{vN}\left(\tilde{\rho}_{pure} \right) - S_{vN}\left(\rho_{Rad} \right),
    \label{srel_is_vNs}
\end{align}
where $\tilde{\rho}_{pure}$ is the a pure state density matrix with the renormalized partition function. Due to \Eqref{srel_is_vNs}, any corrections corrections to the QES formula for the von Neumann entropy will naturally carry over to the relative entropy. In each of our examples, subleading wormhole contributions result in non-perturbatively small corrections to the qJLMS formula at the very least. We demonstrate this in \secref{sec_non_pert}. Larger corrections that are understood to arise for the von Neumann entropy will be the topic of \secref{sec_en_fluc} and \secref{sec_incomp}.

The other natural choice for $\sigma_{code}$ is a pure state that has fidelity $t$ with $\rho_{code}$.\footnote{In general, the fidelity is defined as $F(\rho,\sigma) := \left(\Tr\left[\sqrt{\sqrt{\rho} \sigma \sqrt{\rho}} \right]\right)^2$. However, when either $\rho$ or $\sigma$ is a pure state, this simplifies to an overlap $F(\rho,\sigma) = \Tr\left[\rho \sigma \right]$.} The relative term then becomes
\begin{align}
    \Tr\left[\rho_{Rad} \sigma_{Rad}^{n-1} \right] &= \frac{1}{\left(ke^{\textbf{S}}\right)^n}\left(\sum_{\tau \in S_n, n_1= 1} k^{C\left(\eta^{-1}\circ \tau\right)}\left(e^{\textbf{S}}\right)^{C(\tau)}+ t\sum_{\tau \in S_n, n_1\neq 1} k^{C\left(\eta^{-1}\circ \tau\right)}\left(e^{\textbf{S}}\right)^{C(\tau)}\right).
    \label{pure_gen_rel_term}
\end{align}
This choice leads to qualitatively new corrections of the qJLMS that will be the topic of \secref{sec_rank}.

\subsection{\texorpdfstring{$O\left(e^{-1/G}\right)$}{O(exp(-1/G))}: Wormholes}
\label{sec_non_pert}

The smallest corrections to the qJLMS formula are non-perturbatively small, arising from subleading saddles in the gravitational path integral. These are $O\left(e^{-1/G}\right)$ because the gravitational action is inversely proportional to Newton's constant, $G$. For this purpose, we will only need to consider the black hole in the microcanonical ensemble, where both terms simplify
\begin{align}
    \Tr\left[\rho_{Rad}^n \right] &= \frac{1}{\left(ke^{\textbf{S}}\right)^n}\sum_{\tau \in S_n} k^{C\left(\eta^{-1}\circ \tau\right)}\left(e^{\textbf{S}}\right)^{C(\tau)} \prod_{ i = 1}^{C(\tau)} \Tr\left[\rho_{code}^{n_i} \right],
    \\
    \Tr\left[\rho_{Rad} \sigma_{Rad}^{n-1} \right] &= \frac{1}{\left(ke^{\textbf{S}}\right)^n}\sum_{\tau \in S_n} k^{C\left(\eta^{-1}\circ \tau\right)}\left(e^{\textbf{S}}\right)^{C(\tau)}\Tr\left[\rho_{code} \sigma_{code}^{n_1 - 1} \right]\prod_{ i = 2}^{C(\tau)}\Tr\left[\sigma_{code}^{n_i} \right].
\end{align}
For additional simplicity, we may consider $\rho_{code}$ to be a pure state in the code subspace such that all of its moments are equal to one. Then, we have
\begin{align}
    \Tr\left[\rho_{Rad}^n \right] &= \frac{1}{\left(ke^{\textbf{S}}\right)^n}\sum_{\tau \in S_n} k^{C\left(\eta^{-1}\circ \tau\right)}\left(e^{\textbf{S}}\right)^{C(\tau)} .
\end{align}

Only permutations that maximize the exponents survive the limit where $k$ and $e^{\textbf{S}}$ are sent to infinity. If we keep the ratio $k/e^{\textbf{S}}$ fixed in the limit, these correspond to the \textit{non-crossing permutations} that have $C(\eta^{-1}\circ \tau) + C(\tau) = n+1$ \cite{KREWERAS1972333, SIMION2000367}. The number of non-crossing permutations with $C(\eta^{-1}\circ\tau) = r$ is given by the Narayana number $N_{n,r} := \frac{1}{n}\binom{n}{r}\binom{n}{r-1}$ \cite{KREWERAS1972333, SIMION2000367}, so the sum can be re-expressed in terms of hypergeometric functions
\begin{align}
    \Tr \left[ \rho_{Rad}^n \right] &=\begin{cases}
        k^{1-{n}} \, _2F_1\left(1-{n},-{n};2;\frac{k}{e^{\textbf{S}}}\right), &k < e^{\textbf{S}}
        \\
        (e^{\textbf{S}})^{1-{n}}\, _2F_1\left(1-{n},-{n};2;\frac{e^{\textbf{S}}}{k}\right), &k > e^{\textbf{S}}
    \end{cases} .
\end{align}
Taking the analytic continuation, we find
\begin{align}
    S_{vN}(\rho_{Rad}) = \lim_{n \rightarrow 1} \frac{1}{1-n}{\log\left[\Tr \left[\rho_{Rad}^n \right]\right]}= \begin{cases}
        \log \left[k\right] -\frac{k}{2e^{\textbf{S}}}, & k < e^{\textbf{S}}
        \\
        \log \left[e^{\textbf{S}}\right] -\frac{e^{\textbf{S}}}{2k}, & e^{\textbf{S}} < k
    \end{cases},
\end{align}
which is Page's formula \cite{1993PhRvL..71.1291P}. If we choose $\sigma_{code}$ to be maximally mixed, we may use \Eqref{srel_is_vNs} to find
\begin{align}
\label{srel_exact_mm}
    D(\rho_{Rad} || \sigma_{Rad}) = \begin{cases}
        \frac{k(d_{code} - 1)}{2d_{code}e^{\textbf{S}}}, & k < e^{\textbf{S}}
        \\
        \log \left[\frac{k}{e^{\textbf{S}}}\right] + \frac{e^{2\textbf{S}}d_{code}-k^2}{2 d_{code}e^{\textbf{S}}k},& e^{\textbf{S}}<k <d_{code}e^{\textbf{S}}
        \\
        \log \left[ d_{code}\right] -\frac{e^{\textbf{S}}(d_{code}-1)}{2k}, & d_{code}e^{\textbf{S}} < k
    \end{cases}.
\end{align}

To compare to the qJLMS formula, we need to identify the QES. Prior to the Page time for $\rho_{Rad}$, defined by $\log [k] = \textbf{S}$, the QES and consequently the entanglement wedge of the radiation for any state in the code subspace is the empty set. The area terms in the qJLMS formula, \Eqref{eq:QEJLMS}, are trivially zero while the bulk entropy terms cancel due to $\rho_r$ and $\sigma_r$ being identical, maximally mixed states. On the other hand, the Page time for $\sigma_{Rad}$ occurs when $\log [k] = \log\left[d_{code}\right] +\textbf{S}$. After the Page time for $\rho_R$ but before that for $\sigma_{Rad}$, the QES for $\sigma_{Rad}$ remains the empty set while the QES for $\rho_{Rad}$ becomes nontrivial, located at the black hole horizon. The entanglement wedge for $\rho_R$ thus contains the black hole interior and $\rho_{code}$. The area term for $\rho_R$ is then the black hole entropy while the bulk entropy term disappears because $\rho_{code}$ is pure. Finally, when $\log [k] > \log\left[d_{code}\right] +\textbf{S}$, the QESs for both $\rho_{Rad}$ and $\sigma_{Rad}$ are nontrivial, with areas canceling in \Eqref{eq:QEJLMS}. The bulk term for $\sigma_{code}$ is $\log\left[ d_{code}\right]$. In total, the qJLMS formula gives
\begin{align}
    D_{qJLMS}(\rho_{Rad} || \sigma_{Rad}) = \begin{cases}
        0, & k < e^{\textbf{S}}
        \\
        \log \left[\frac{k}{e^{\textbf{S}}}\right] ,& e^{\textbf{S}}<k <d_{code}e^{\textbf{S}}
        \\
        \log \left[ d_{code}\right] , & d_{code}e^{\textbf{S}} < k
    \end{cases},
    \label{micro_qjlms_mm}
\end{align}
which is only different from \Eqref{srel_exact_mm} by terms that are non-perturbatively small. These corrections become $O(1)$ around the transitions between the three regimes.

We may also take $\sigma_{code}$ to be a pure state which, in general, has a fidelity of $t$ with the pure state $\rho_{code}$.
The $t = 0$ case was the subject of Refs.~\cite{2021PhRvL.126q1603K,2021PRXQ....2d0340K} where it was noted that the dominant permutations for the first term of \Eqref{pure_gen_rel_term} are those that are the identity on the first element and non-crossing on the rest. There are now $N_{n-1,r}$ such non-crossing permutations with $C(\eta^{-1}\circ\tau) = r$, so the first sum may be written once more as a hypergeometric function. The second sum consists of all non-crossing permutations that are left over, so there are $N_{n,r} - N_{n-1,r}$ of these. We have
\begin{align}
    \Tr\left[\rho_{Rad} \sigma_{Rad}^{n-1} \right]\hspace{-.1cm} &=\hspace{-.1cm} \begin{cases} k^{1-n}\left((1-t){}_2F_1\left(1-n,2-n;2;\frac{k}{e^{\textbf{S}}} \right)+t {}_2F_1\left(1-n,-n;2;\frac{k}{e^{\textbf{S}}} \right)\right), & \hspace{-.26cm}k < e^{\textbf{S}}
    \\
     e^{\textbf{S}(1-n)}\left((1-t){}_2F_1\left(1-n,2-n;2;\frac{e^{\textbf{S}}} {k}\right)+t {}_2F_1\left(1-n,-n;2;\frac{e^{\textbf{S}}}{k} \right)\right), & \hspace{-.26cm} e^{\textbf{S}} < k
    \end{cases}\hspace{-.15cm}.
\end{align}
Taking the replica limit, we find
\begin{align}
    D(\rho_{Rad} || \sigma_{Rad}) = \begin{cases}
        (1-t)\left(1+\frac{k}{2e^{\textbf{S}}}+\left(\frac{e^{\textbf{S}}}{k}-1\right) \log
   \left[1-\frac{k}{e^{\textbf{S}}}\right]\right),  & k < e^{\textbf{S}}
   \\
   \infty, & e^{\textbf{S}}<k
    \end{cases},
\end{align}
where in the second line, we have assumed that $t < 1$. If $t = 1$, the states are identical and the relative entropy is always zero. Because $\rho_{code}$ and $\sigma_{code}$ are both pure, the QESs for $\rho_{Rad}$ and $\sigma_{Rad}$ are identical in all regimes. Before the Page time, these are the empty set, leading to trivial relative entropy. After the Page time, they are nontrivial and the infinite relative entropy between $\rho_{code}$ and $\sigma_{code}$ lead to infinite relative entropy in the radiation
\begin{align}
    D_{qJLMS}(\rho_{Rad} || \sigma_{Rad}) = \begin{cases}
    0,  & k < e^{\textbf{S}}
   \\
   \infty, & e^{\textbf{S}}<k
    \end{cases}.
    \label{pure_micro_qjlms}
\end{align}
Again, the corrections from the exact formula are non-perturbatively small up until the Page time where they become $O(1)$.

We note that all results from this section also apply to AdS/CFT in higher dimensions in so-called \textit{fixed-area states} \cite{2019JHEP...10..240D,2019JHEP...05..052A,2020JHEP...03..191D} where two extremal surfaces in the bulk are fixed in the gravitational path integral to have definite area. In those examples, ${Rad}$ is interpreted as a subregion of the boundary and the bulk state is the state of the quantum fields located between the two fixed area surfaces.

\subsection{\texorpdfstring{$O\left(G^{-1/2}\right)$}{O(1/sqrt(G))}: Energy Fluctuations}
\label{sec_en_fluc}

New phenomena arise when we allow the the energy of the black hole to fluctuate in the canonical ensemble. Indeed, energy fluctuations in condensed matter systems were first understood to give rise to $O(\sqrt{\mbox{Volume}})$ corrections to volume-law entanglement near entanglement phase transitions \cite{2017PhRvL.119v0603V,2019PhRvE.100b2131M}. In order to observe analogous corrections to qJLMS, we consider $\sigma_{code}$ in the maximally mixed state and a sufficiently large code subspace to find a separation of time scales. We can then use \Eqref{srel_is_vNs} such that all we need is to evaluate von Neumann entropies. For the same reasons as \Eqref{micro_qjlms_mm}, the qJLMS predicts 
\begin{align}
    D_{qJLMS}(\rho_{Rad} || \sigma_{Rad}) = \begin{cases}
        0, & k < e^{{S}_{BH}}
        \\
        \log \left[\frac{k}{e^{{S}_{BH}}}\right] ,& e^{{S}_{BH}}<k <d_{code}e^{{S}_{BH}}
        \\
        \log \left[ d_{code}\right] , & d_{code}e^{{S}_{BH}} < k
    \end{cases}.
    \label{can_qjlms_mm}
\end{align}

In the canonical ensemble, it is not easy to explicitly evaluate the sum over permutations and then analytically continue $n$ to one to find the von Neumann entropy. Instead, one may solve for the entanglement spectrum, $D(\lambda)$, explicitly, then evaluate the von Neumann entropy as
\begin{align}
    S_{vN}(\rho_{Rad}) = -\int d\lambda D(\lambda) \lambda \log\left[ \lambda\right]
\end{align}
The entanglement spectrum can be extracted from the resolvent
\begin{align}
    D(\lambda) = -\frac{1}{\pi}\lim_{\epsilon\rightarrow 0}\Im \left[R(\lambda+i\epsilon)\right], \quad R(\lambda) = \Tr\left[ \frac{1}{\lambda- \rho_{Rad}}\right].
\end{align}
An implicit formula for the resolvent for the canonical ensemble was determined in Ref.~\cite{2019arXiv191111977P} (and reviewed in Appendix~\ref{app_resolvent}) using a Schwinger-Dyson equation
\begin{align}
    \lambda R(\lambda) = k + \int_0^{\infty}  ds \frac{\rho(s)w(s) R(\lambda)}{k-w(s) R(\lambda)}, \quad w(s) := \frac{y(s)}{Z_1}.
\end{align}
There is no closed-form solution to this equation but an approximation to the spectrum that works to high accuracy was found to be a \textit{shifted cutoff thermal spectrum}
\begin{align}
    \label{approx_spec}
    D(\lambda) = \int_0^{s_k} ds \rho(s)\delta(\lambda- \lambda_0-w(s))
,\quad
    k = \int_0^{s_k}ds \rho(s),\quad \lambda_0 = \frac{1}{k}\int_{s_k}^{\infty}\rho(s) w(s).
\end{align}
As the name suggests, this is the thermal spectrum of JT gravity shifted by $\lambda_0$ with a cutoff after the first $k$ eigenvalues.

Before the Page time, both density matrices in \Eqref{srel_is_vNs} are nearly maximally mixed and the relative entropy is consequently $O\left(\frac{k}{e^{S_{BH}}}\right)$, the familiar non-perturbative corrections to qJLMS found in \secref{sec_non_pert}. More interestingly, in a window of size $\Delta \log k \sim O(G^{-1/2})$ around the Page time, there is a \textit{negative} $O(G^{-1/2})$ correction to the naive Page curve of $\min\left( \log[k], S_{BH}\right)$ \cite{2019arXiv191111977P}. In \Eqref{srel_is_vNs}, this only occurs for the second term because the first term has yet to reach the Page time. This leads to $D(\rho_{Rad}^{(a)}||\sigma_{Rad})$ being $O\left(G^{-1/2}\right)$ and large in the semi-classical limit. This represents a significant breakdown of the qJLMS formula.

After the Page time, the entropy of $\rho_{Rad}$ becomes exponentially close to $S_{BH}$, while the entropy of $\tilde{\rho}_{pure}$ remains exponentially close to $\log \left[k\right]$, such that the relative entropy is large.

\begin{figure}
    \centering
    \includegraphics[width = .6\textwidth]{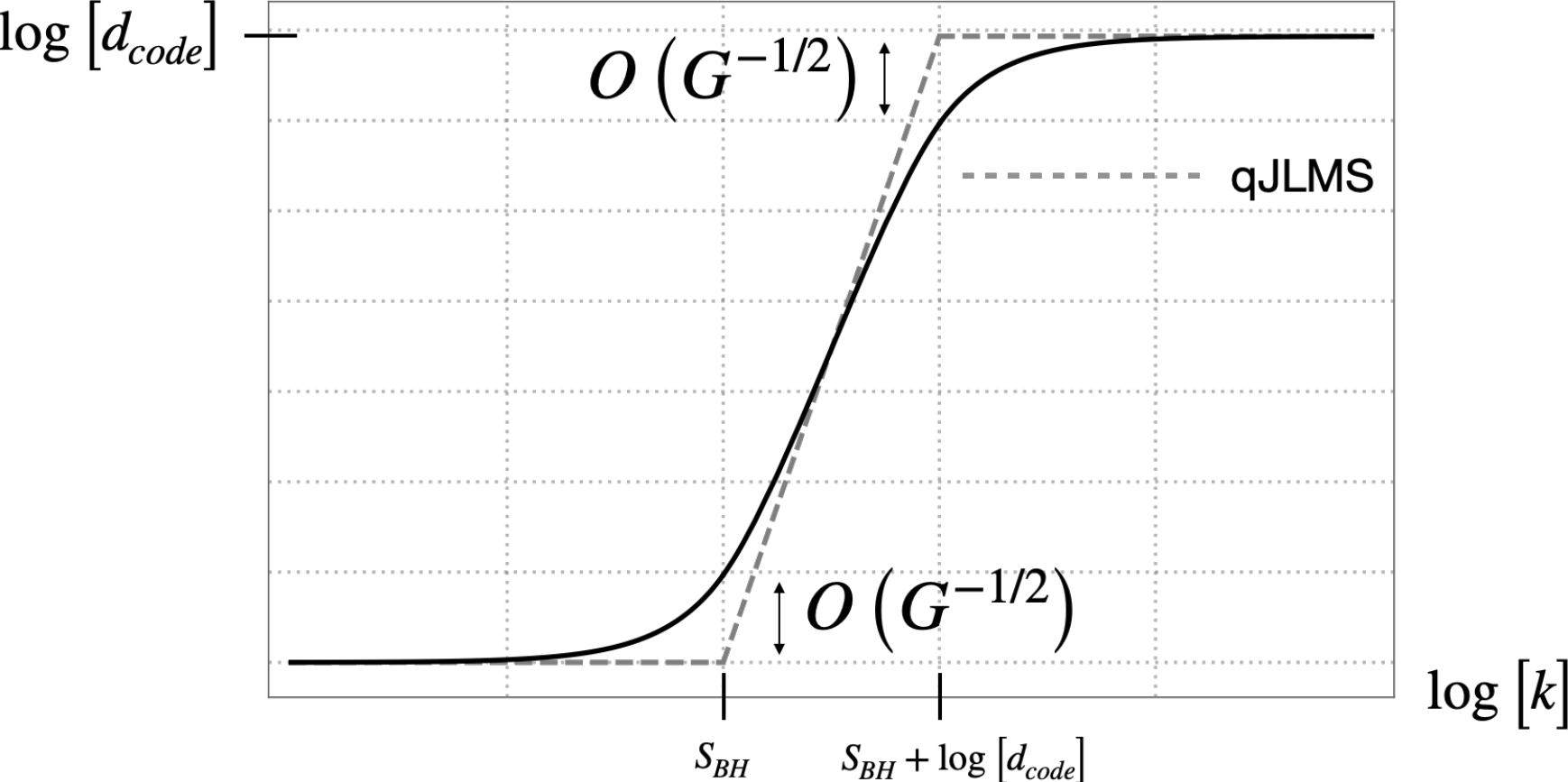}
    \caption{A sketch of the relative entropy (black line) for the canonical ensemble with $\sigma_{code}$ being the maximally mixed state in the code subspace. There are two locations with $O\left(G^{-1/2}\right)$ corrections.}
    \label{energy_fluctuation_jlms_fig}
\end{figure}

The qJLMS formula fails once again when $\tilde{\rho}_{pure}$ undergoes its Page transition at $\log[k] = S_{BH}+ \log\left[d_{code}\right]$. At this point, the entropy of $\rho_{Rad}$ is still exponentially close to $S_{BH}$, but the entropy of $\tilde{\rho}_{pure}$ receives an $O\left(G^{-1/2}\right)$ correction. 

After this transition, the entropies are exponentially close to $S_{BH}$ and $S_{BH} + \log\left[d_{code}\right]$ respectively and the corrections to qJLMS are, once more, non-perturbatively small. All together, we have (see \figref{energy_fluctuation_jlms_fig})
\begin{align}
    D(\rho_R || \sigma_R) = \begin{cases}
    O\left(\frac{k}{e^{S_{BH}}}\right), & k \ll e^{S_{BH}}
    \\
    O\left(\left(\beta G\right)^{-1/2} \right), &  k \simeq e^{S_{BH}}
    \\
    \log\left[\frac{k}{e^{{S}_{BH}}}\right] + O\left(\frac{e^{S_{BH}}}{k} , \frac{k}{d_{code}e^{S_{BH}}}\right), & e^{S_{BH}}\ll k \ll d_{code}e^{S_{BH}}
    \\
    \log\left[ d_{code}\right] - O\left(\left(\beta G\right)^{-1/2} \right), & k \simeq d_{code}e^{S_{BH}}
    \\
    \log\left[ d_{code}\right] + O\left(\frac{d_{code}e^{S_{BH}}}{k}\right), & k \simeq d_{code}e^{S_{BH}}
    \end{cases}.
    \label{srel_latetime_canon}
\end{align}

In higher dimensional AdS/CFT, analogous $O(G^{-1/2})$ corrections to the von Neumann entropy have been evaluated near entanglement phase transitions \cite{2020JHEP...12..084M,2020JHEP...11..007D}. These have been attributed to fluctuations in the \textit{areas} of the QESs, which are directly related to energy fluctuations in the canonical ensemble. We expect the $O(G^{-1/2})$ corrections to qJLMS to persist more generally and this will be made more precise in \secref{sec:discuss}.

\subsection{\texorpdfstring{$O\left(G^{-1}\right)$}{O(1/G)}: Incompressibility}
\label{sec_incomp}

So far, we have mainly been concerned with pure states for $\rho_{code}$. Pure states fall under the umbrella of \textit{perfectly compressible} quantum states, those that may be well-approximated by keeping only $e^{S(\rho_{code})}$ of the states in their support. Interestingly, it was argued in Ref.~\cite{2021JHEP...04..062A} that incompressible bulk states lead to leading order ($O(G^{-1})$) corrections to the QES formula. By \Eqref{srel_is_vNs}, these states lead to $O(G^{-1})$ corrections of the qJLMS formula. As an example of an incompressible state, we now take $\rho_{code}$ to be a probabilistic mixture of a pure state and the maximally mixed state:
\begin{align}
\rho_{code} = p \ket{\psi}\bra{\psi} + (1-p) \frac{\mathbbm{1}_{code}}{d_{code}}.
\end{align} 
The eigenvalues and hence, the entropy of $\rho_{code}$ is easily computable. For large code subspaces, we have $S(\rho_{code}) = (1-p) \log\left[{d_{code}}\right] + O(d_{code}^{0})$.

\begin{figure}
    \centering
    \includegraphics[width = .6\textwidth]{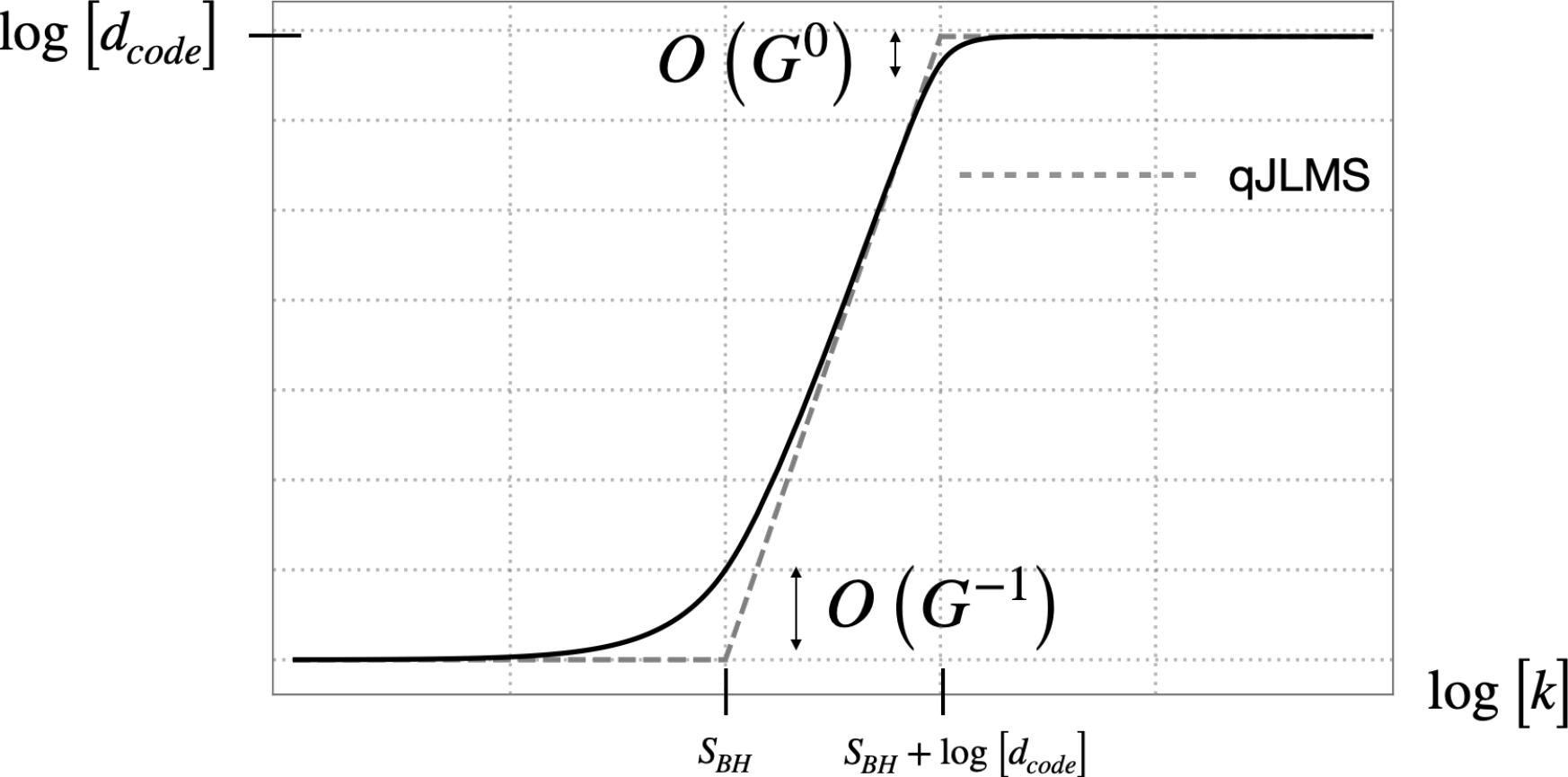}
    \caption{A sketch of the relative entropy (black line) for the microcanonical ensemble with $\sigma_{code}$ being the maximally mixed state in the code subspace and $\rho_{code}$ an incompressible state. The large $O\left(G^{-1}\right)$ correction only occurs at the first transition.}
    \label{incompressible_jlms_fig}
\end{figure}

The boundary entropy in such a mixture was computed for fixed-area states (equivalently the microcanonical ensemble) \cite{2021JHEP...04..062A}
\begin{align}
	{S(\rho_{Rad}) = \begin{cases}
	\log [k], & k \ll \frac{e^{\textbf{S}}}{p}\\
	p \textbf{S} + (1-p) \log [k], &\frac{e^{\textbf{S}}}{p}\ll k \ll e^{\textbf{S}+S(\rho_{code})}\\
	\textbf{S} + S(\rho_{code}), & e^{\textbf{S}+S(\rho_{code})} \ll k
	\end{cases}},
\end{align}
where we have dropped $O(1)$ and more subleading corrections. At this order, we thus obtain 
\begin{align}
    {D(\rho_{Rad} || \sigma_{Rad}) = 
    \begin{cases}
	0, & k \ll \frac{e^{\textbf{S}}}{p}\\
	p(\log[k]- \textbf{S}), &\frac{e^{\textbf{S}}}{p}\ll k \ll e^{\textbf{S}+S(\rho_{code})}\\
	\log[k]-\textbf{S}-S(\rho_{code}),&e^{\textbf{S}+S(\rho_{code})}\ll k \ll d_{code}e^\textbf{S}
	\\
	\log[d_{code}]- S(\rho_{code}), & d_{code}e^\textbf{S} \ll k
	\end{cases}},
\end{align}
which involves $O(G^{-1})$ corrections to the qJLMS answer (see \figref{incompressible_jlms_fig})
\begin{align}
    {D(\rho_{Rad} || \sigma_{Rad}) = 
    \begin{cases}
	0, & k \leq e^{\textbf{S}+S(\rho_{code})}\\
	\log[k]-\textbf{S}-S(\rho_{code}), & e^{\textbf{S}+S(\rho_{code})}\leq k \leq d_{code}e^\textbf{S} 
	\\
	\log[d_{code}]- S(\rho_{code}), & d_{code}e^\textbf{S} \leq k
	\end{cases}}.
\end{align}

\subsection{Infinite: Rank Deficiency}
\label{sec_rank}

In the canonical ensemble, we can also take $\rho_{code}$ and $\sigma_{code}$ to be pure states with fidelity $t$. The description of the QESs is identical to the microcanonical case (\Eqref{pure_micro_qjlms}) with the microcanonical black hole entropy replace by the canonical $S_{BH}$
\begin{align}
    D_{qJLMS}(\rho_{Rad} || \sigma_{Rad}) = \begin{cases}
    0,  & k < e^{S_{BH}}
   \\
   \infty, & e^{S_{BH}}<k
    \end{cases}.
    \label{pure_can_qjlms}
\end{align}
Summing all terms, we have
\begin{align}
    \Tr\left[\rho_{Rad} \sigma_{Rad}^{n-1} \right] &= \frac{1}{\left(kZ_1\right)^n}\sum_{\tau \in S_n, n_1= 1} k^{C\left(\eta^{-1}\circ \tau\right)}\prod_{ i = 1}^{C(\tau)} Z_{n_i} 
    +
    \frac{t}{\left(kZ_1\right)^n}\sum_{\tau \in S_n, n_1\neq 1} k^{C\left(\eta^{-1}\circ \tau\right)}\prod_{ i = 1}^{C(\tau)} Z_{n_i} 
    \nonumber
    \\
    &= (1-t)\frac{\Tr\left[\rho_{Rad}^{n-1} \right]}{k}+ t\Tr\left[\rho_{Rad}^{n} \right].
\end{align}
We need to take care of the replica limit
\begin{align}
    \lim_{n\rightarrow 1} \frac{1}{1-n} \log\left[\Tr\left[\rho_{Rad}\sigma_{Rad}^{n-1} \right] \right] 
    &=-\frac{1-t}{k}{\partial_n \Tr\left[\rho_{Rad}^{n-1} \right]\big|_{n = 1}} - t\partial_n\Tr\left[\rho_{Rad}^{n} \right]\big|_{n = 1}.
\end{align}
Re-expressing the traces as integrals over the entanglement spectrum of $\rho_{Rad}$, we have
\begin{align}
        \lim_{n\rightarrow 1} \frac{1}{1-n} \log\left[\Tr\left[\rho_{Rad}\sigma_{Rad}^{n-1} \right] \right] &= -\frac{1-t}{k}\partial_n \int d\lambda D(\lambda) {\lambda^{n - 1}}\big|_{n = 1}-t\partial_n \int d\lambda D(\lambda) {\lambda^{n }}\big|_{n = 1}
        \nonumber
        \\
        &= -\frac{1-t}{k}\int d\lambda D(\lambda){\log[\lambda]}-t\int d\lambda D(\lambda)\lambda{\log[\lambda]}.
        \label{relative_eig_distrib}
\end{align}

In total, the relative entropy is thus
\begin{align}
    D\left(\rho_{Rad}||\sigma_{Rad}\right)= -(1-t)\int d\lambda D(\lambda) \left(\frac{1}{k} - \lambda\right){\log[\lambda]}.
    \label{srel_canon_pure}
\end{align}
Prior to the Page time, the entanglement spectrum is a sharply peaked semi-circle of eigenvalues centered around $\frac{1}{k}$ due to the dominance of the identity permutation in the sum. Thus, we see that the relative entropy is close to zero, with the non-perturbative corrections to the qJLMS formula arising strictly from the finite width of the spectrum. 

\begin{figure}
    \centering
    \includegraphics[width = .6\textwidth]{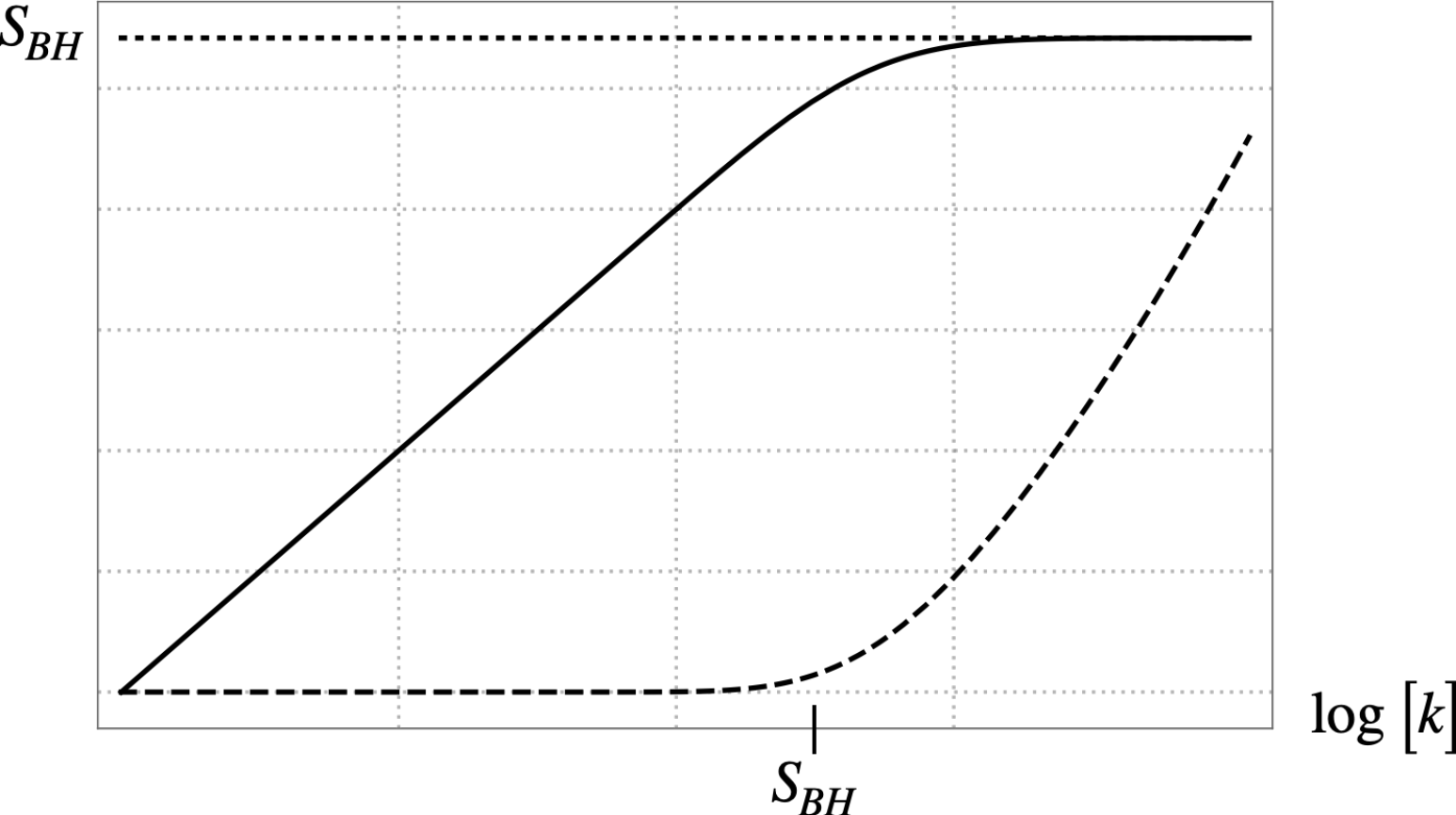}
    \includegraphics[width=\textwidth]{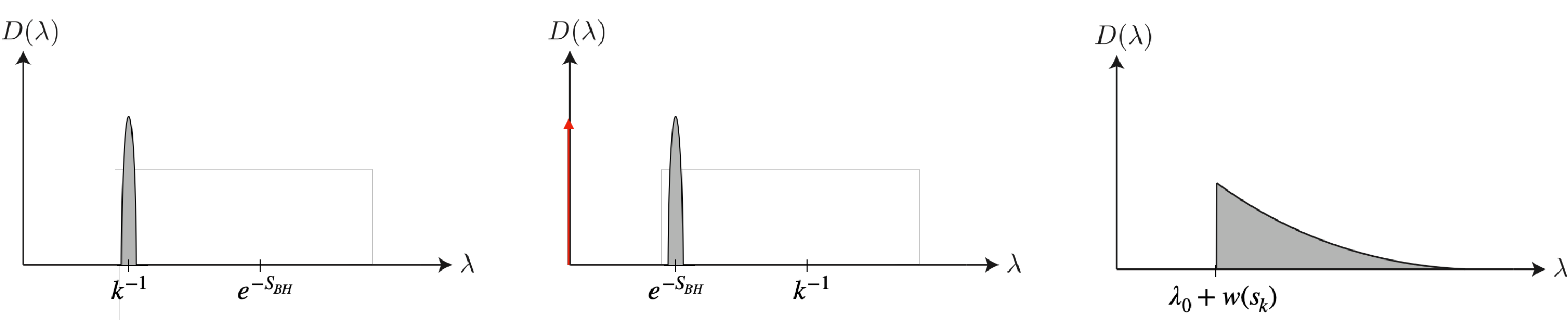}
    \caption{Top: The relative entropy between pure states in the code subspace (dashed line) is shown alongside the von Neumann entropy (solid line) and black hole entropy (dotted line) using the approximate entanglement spectrum \Eqref{approx_spec}. Unlike the microcanonical ensemble, the relative entropy is smooth and finite across the Page transition. It grows unboundedly as $(\log k)^2$ at large $k$. Bottom: We sketch the spectrum of $\rho_{Rad}$ prior to the Page time (left) after the Page time in the microcanonical ensemble (middle) and after the Page time in the canonical ensemble (right). The red arrow indicates an delta function at zero.}
    \label{srel_canon_pure_approx}
\end{figure}

While, as before, there will be large corrections near the phase transition,\footnote{In fact, we expect these to be $O\left(G^{-1}\right)$ as in the logarithmic negativity \cite{2021arXiv211011947D} because the integral is most sensitive to the smallest eigenvalues. This behaviour has already been seen in the Renyi entropies for $n<1$ in Refs.~\cite{2021arXiv211011947D,2022arXiv220111730A}.} we focus on a more striking feature of \Eqref{srel_canon_pure}, namely that after the Page transition, the ``corrections'' to the qJLMS formula are \textit{infinite}. The qJLMS formula after the Page time predicts an infinite relative entropy because the states in the code subspace are pure. From \Eqref{srel_canon_pure}, we see that the relative entropy is finite as long as there are no eigenvalues at zero. Indeed, this is the case, as carefully examined in Ref.~\cite{2019arXiv191111977P} and reviewed in \secref{sec_en_fluc}. The entanglement spectrum is a ``shifted cutoff thermal spectrum'' where the smallest eigenvalue lies a finite distance away from zero, as sketched in \figref{srel_canon_pure_approx}. This is reasonable because JT gravity has an infinite number of eigenstates.\footnote{For this phenomenon to occur, one only needs the number of accessible eigenstates in the black hole spectrum to be much larger than $e^{S_{BH}}$ and not necessarily infinite. These violations will occur for $k$ larger than $e^{S_{BH}}$ but smaller than the number of accessible eigenstates.} Seemingly innocuous small tails in the spectrum at high energies drastically change the relative entropy. 

Note that in the microcanonical ensemble (or equivalently random tensors), there is only a finite number of states accessible in JT gravity, leading to a rank deficiency (zero eigenvalues) following the Page time and an infinite relative entropy.

While the upshot is that there is an infinite violation of the qJLMS formula, to get a sense of the true magnitude of the relative entropy, we can approximate the entanglement spectrum using the shifted cutoff thermal spectrum.
Assuming the validity of this approximation in the computation of relative entropy,
\begin{align}
    D\left(\rho_{Rad}||\sigma_{Rad}\right)= -\int_0^{s_k} ds \rho(s) \left(\frac{1}{k} - \lambda_0 - w(s)\right){\log[\lambda_0 +w(s)]},
\end{align}
which we plot in \figref{srel_canon_pure_approx} alongside the Page curve. While the relative entropy is finite, unlike \Eqref{srel_latetime_canon}, it is unbounded at large $k$ and grows asymptotically as $(\log k)^2$.\footnote{When taking the strict $G\rightarrow 0$ limit as considered in Refs.~\cite{2021arXiv211212828W, 2021arXiv211212156L, 2021arXiv211005497L}, the above violation of the JLMS formula formally disappears since $\log k \rightarrow \infty$, and both bulk and boundary relative entropies are infinite. Our violations are nevertheless arbitrarily large when considering $G$ to be small but finite. We thank Geoff Penington for discussions related to this.}

It is instructive to analyze the relative entropy on the black hole system rather than the radiation. Indeed, it was this complementary relative entropy that was important in the exact reconstruction analysis of Ref.~\cite{2016PhRvL.117b1601D}. The entropy term is identical to that for the radiation because the global state is pure. The relative term does not have this symmetry and instead reads
\begin{align}
    \Tr\left[\rho_B \sigma_B^{n-1} \right] = \frac{1}{\left(kZ_1\right)^n}\sum_{\tau \in S_n, \tilde{n}_1 = 1} k^{C\left(\tau\right)}\prod_{ i = 1}^{C(\eta^{-1}\circ\tau)} Z_{\tilde{n}_i}
    +
    \frac{t}{\left(kZ_1\right)^n}\sum_{\tau \in S_n, \tilde{n}_1 \neq 1} k^{C\left(\tau\right)}\prod_{ i = 1}^{C(\eta^{-1}\circ\tau)} Z_{\tilde{n}_i},
\end{align}
where $\tilde{n}_i$ is the length of the $i^{th}$ cycle of $\eta^{-1}\circ \tau$. After the Page time, the entanglement wedge of the black hole does not include the interior because the interior is part of the entanglement wedge of the radiation. Therefore the qJLMS formula predicts zero relative entropy for the black hole. In this regime, the identity permutation dominates the sum, leading to 
\begin{align}
    \Tr\left[\rho_B \sigma_B^{n-1} \right] \simeq \frac{Z_n}{Z_1^n} \simeq \Tr\left[ \rho_B^n\right].
\end{align}
Therefore, the true relative entropy is indeed close to zero. The leading corrections to zero relative entropy arise from the $\binom{n}{2}$ replica wormholes that involve a single two-boundary wormhole and $(n-2)$ single-boundary spacetimes. This will be $O\left(\frac{e^{S_{BH}}}{k} \right)$, so we conclude that even when there is an infinite violation of the qJLMS formula for the radiation system, there is only a non-perturbatively small correction to the qJLMS formula for the black hole. This makes clear that reconstruction of interior operators from the radiation succeeds with high fidelity. This is manifest from calculations of the fidelity of the Petz map in Ref.~\cite{2019arXiv191111977P, 2021arXiv211200020V}. 

\section{Summary and Discussion}

\label{sec:discuss}

In this paper, we have studied corrections and violations of the qJLMS formula. While some corrections are very small (non-perturbatively so), they play important roles, such as in the distinguishability of black hole microstates \cite{2021PhRvL.126q1603K, 2021PRXQ....2d0340K} and approximate quantum error correction \cite{2019JHEP...12..007H,2021arXiv211212789G}. We found larger corrections to the qJLMS formula at $O\left(G^{-1/2}\right)$ and $O\left(G^{-1}\right)$ that imply breakdowns of entanglement wedge reconstruction near entanglement phase transitions. Finally, we found infinite violations of the qJLMS formula in the PSSY model, our most curious result. We conclude with a few comments on open questions.

\paragraph{Single Instances and Ensemble Averages}
Famously, JT gravity is dual to an ensemble of quantum mechanical systems rather than an individual theory \cite{2019arXiv190311115S}. It is then natural to ask whether the corrections we find hold for single instances of the ensemble. For the infinite violations, this can be confirmed immediately because if the ensemble average of the relative entropy is finite, so is the relative entropy for a measure one set of single instances of the ensemble. For finite corrections, one needs to evaluate the variance in the relative entropy which may be seen to be small relative to the mean by a consideration of the genus number of the leading saddle point configurations in the replica trick (see e.g.~Ref.~\cite{2021PRXQ....2d0340K}).

Our bulk calculations were performed in the no-boundary state. It may be interesting to perform bulk computations in single draws of the ensemble by using $\alpha$-states instead \cite{2020JHEP...08..044M, 2021arXiv210316754S,2022arXiv220307384B}.

\paragraph{Diagonal approximation}

As we have previously noted, all of our calculations in the microcanonical ensemble are directly transferable to fixed-area states in higher dimensional AdS/CFT. While fixed-area states are interesting in their own right, it is desirable to understand generic, unconstrained holographic states, such as the vacuum or thermal state. 

Fixed-area states provide a basis in which smooth geometric states can be decomposed. For example, for a state with two candidate RT surfaces for a given subregion $R$, we have
\begin{align}
    \ket{\psi} = \sum_{A_1, A_2}\sqrt{{P(A_1,A_2)}}\ket{\psi}_{A_1, A_2},
\end{align}
where $P(A_1, A_2)$ is a probability density and is, semi-classically, a Gaussian with $O\left(G^{1/2}\right)$ width \cite{2020JHEP...12..084M}.
The reduced density matrix on $R$ is thus given by
\begin{align}
    \rho_R = \sum_{A_1,A_1' A_2,A_2'}\sqrt{{P(A_1,A_2)P(A_1',A_2')}}\Tr_{B}\left[ {\ket{\psi}_{A_1,A_2}\bra{\psi}_{A_1',A_2'}}\right].
\end{align}
As in Ref.~\cite{2020JHEP...12..084M}, we can first assume that $\rho_A(A_1,A_2)$'s with different areas live in orthogonal subspaces 
\begin{align}
    \rho_R = \bigoplus_{A_1,A_2} P(A_1, A_2)\rho_A(A_1,A_2)+\mbox{OD},
    \label{direct_sum_rhoA}
\end{align}
where OD represents the ``off-diagonal'' terms in the trace for which $(A_1, A_2) \neq (A_1', A_2')$,
and likewise for $\sigma_R$. If we can drop the off-diagonal terms in a ``diagonal approximation,'' \cite{2020JHEP...12..084M,2020JHEP...11..007D} the relative entropy becomes
\begin{align}
    D(\rho_R  || \sigma_R) = \sum_{A_1, A_2} P(A_1,A_2) D(\rho_R(A_1,A_2)||\sigma_R(A_1,A_2)).
\end{align}
$O\left(G^{-1/2}\right)$ corrections will arise for generic $\sigma_A$ from the sum over $A_1$ and $A_2$. Intuitively, we expect that such off-diagonal terms should decrease the relative entropy because removing them corresponds to applying a dephasing channel to the states prior to tracing over the complement of $R$ and relative entropies are monotonically decreasing under quantum channels.

In fact, in the particular case where $\sigma_R$ is maximally mixed in the code subspace
\begin{align}
    \sigma_R = \sum_{A_1,A_1' A_2,A_2'}\sqrt{{P(A_1,A_2)P(A_1',A_2')}}\Tr_{\bar{R}}\left[\frac{1}{d_{b}}\sum_{\psi} \frac{\ket{\psi}_{A_1,A_2}\bra{\psi}_{A_1',A_2'}}{\bra{\psi}\psi\rangle}\right],
\end{align}
we can make a precise argument for the diagonal approximation. The diagonal approximation for the entropy term in the relative entropy has already been understood for general states in Ref.~\cite{2021JHEP...04..062A}. It was shown that up to an $O(\log [G])$ discrepancy, the entropy of $\rho_R$ agrees with the entropy computed using the diagonal approximation. The relative term, $\Tr\left[\rho_R \sigma_R^{n-1} \right]$, can also be written as an entropy-like term, $\Tr\left[\rho_R^n\right]$, by renormalizing $A_2$ to $A_2 + \log\left[d_{b}\right]$, just as in \Eqref{srel_is_vNs}. The relative entropy is given by a difference of two von Neumann entropies, each of which are at most $O(\log [G])$ different than their diagonal approximations. Of course, $\log [G] \ll G^{-1/2}, G^{-1}$ in the semi-classical limit, so all large corrections to the von Neumann entropy near phase transitions \cite{2020JHEP...12..084M,2020JHEP...11..007D,2021JHEP...04..062A} lead to corrections of the relative entropy at the same order. Thus, we see that for a maximally mixed state in the code subspace, there will be large corrections in higher dimensions, just as we found in the PSSY model. We expect this to be more generally true although we leave such an argument for future work.

\paragraph{Infinite violations in higher dimensions?} The PSSY model is very simple and quite special. It is important to understand if the infinite violations of the qJLMS formula that we found in \secref{sec_rank} appear more generically in AdS/CFT, namely in higher-dimensions. If we are working with generic (not fixed-area) states in AdS/CFT, we expect that there are not rank deficiencies in the boundary CFT that lead to infinite relative entropies for an open region $R$, assuming the vector ``purifying $\sigma_R$'' is separating, which is the case for the vacuum sector by the Reeh-Schleider theorem (see e.g.~\cite{2018arXiv180304993W}). We can then expect similar infinite violations of the qJLMS formula if the bulk states lying between two extremal surfaces have infinite relative entropy. 

It is common in the literature to treat the code subspace in the bulk to be factorizable into Hilbert spaces (and Type I von Neumann algebras) associated with each boundary region, which would certainly lead to such infinite violations. Nevertheless, the code subspace should accurately be treated as a (non-factorizable) quantum field theory where subregions are described not by Hilbert spaces, but Type III${}_1$ algebras. Then, we expect the bulk relative entropy to be finite as well if the bulk state is separating. 

However, there may be physically relevant QFT states that aren't separating.\footnote{We thank Tom Faulkner for discussions on this topic.} A possible example is that of Connes-cocyle flowed states which have not been proved to be cyclic and separating \cite{connes1973classification,2018arXiv181204683C}. Such states do have an interesting semiclassical bulk description \cite{2020PhRvD.101d6001B, 2020PhRvD.102f6008B} and thus, it would be interesting to understand if they lead to infinite bulk relative entropies and thus, potentially similar violations of the qJLMS formula.

\paragraph{Infinite violations in Tensor Networks} As discussed in \secref{PSSY_sec}, the states in the PSSY model can be given a tensor network description. These tensor networks have the interesting feature that they contain non-maximally entangled bonds. The important feature that led to the infinite violations of the qJLMS formula in the PSSY model was that while the entropy of the black hole was smaller than the radiation, the rank of the thermal spectrum was much larger. One can construct tensor networks with such non-maximally entangled bonds with the above feature and we expect to find infinite violations of the qJLMS formula in these cases as well.

\paragraph{Operator version of JLMS} 
 In our analysis, we found large corrections to the expected holographic relative entropy formula. However, another result in Ref.~\cite{2016JHEP...06..004J} that we refer to as the operator JLMS formula is
\begin{equation}
    \hat{K}_R = \frac{\hat{\mathcal{A}}^{\gamma_{\rho}}}{4G} + \hat{K}_r,
\end{equation}  
where $K_R=-\log \rho_R$ is the boundary modular Hamiltonian, $K_r$ is the bulk modular Hamiltonian in the entanglement wedge of $R$ and $\gamma_{\rho}$ is the QES. 

It is easy to see that our large corrections in fact arise from large corrections to the operator JLMS formula due to the sensitivity of the logarithm to small eigenvalues. As a simple example, consider two density matrices
\begin{align}
    &\rho_1=\begin{bmatrix}
1 & 0 \\
0 & 0
\end{bmatrix},
&\rho_2 = \begin{bmatrix}
1-\epsilon & 0 \\
0 & \epsilon
\end{bmatrix}.
\end{align}
$\rho_1$ and $\rho_2$ are close in trace distance, but the corresponding modular Hamiltonians $K_1$ and $K_2$ aren't. More generally, the relative entropy satisfies a continuity property as long as the eigenvalues remain non-zero \cite{2005JMP....46j2104A}.

\paragraph{R\'enyi relative entropies} Just as the classical Kullback-Leibler divergence can be generalized to $\alpha$-R\'enyi divergences \cite{10020820209}, the quantum relative entropy can be generalized to R\'enyi relative entropies, such as the Petz R\'enyi relative entropy \cite{1986RpMP...23...57P} or the Sandwiched R\'enyi relative entropy \cite{2013JMP....54l2203M,2014CMaPh.331..593W}. These quantities are of great interest as they have complementary operational meanings and have the potential to teach us more about bulk reconstruction.\footnote{In the exact error-correction limit, they obey R\'enyi JLMS formulas \cite{2018JHEP...10..036M}.} They may be evaluated using replica tricks, as was exploited in the case where the bulk states were orthogonal in the code subspace in Ref.~\cite{2021PRXQ....2d0340K}, and we expect this can be generalized to the states studied in this paper. One interesting feature of the R\'enyi relative entropies is that they are finite when $\alpha$ is less than one, regardless of the supports of the density matrices, so the divergences we see may be under better control. 
\acknowledgments

We would like to thank Reginald Caginalp and Mudassir Moosa for initial collaboration on this project and Vladimir Narovlansky and Shinsei Ryu for previous collaboration. We thank Chris Akers, Xi Dong, Tom Faulkner, and Geoff Penington for helpful discussions and Chris Akers, Don Marolf, and Vladimir Narovlansky for comments on the draft.
JKF is supported through a Simons Investigator Award to Shinsei Ryu from the Simons Foundation (Award Number: 566166). PR is supported in part by a grant from the Simons Foundation, and by funds from UCSB. This material is based upon work supported by the Air Force Office of Scientific Research under award number FA9550-19-1-0360. 
 
\appendix

\section{Resolvent for the Relative Entropy}
\label{app_resolvent}
In this appendix, we describe a resolvent method for computing the relative entropy. This is an alternate way to derive the results we found in \secref{sec_violations}. In analogy with the resolvent trick for other quantum information quantities \cite{2019arXiv191111977P,2021PRXQ....2c0347S,2021arXiv211209122A}, one expects this to be a useful technique for performing analytic continuation.

First we briefly review the resolvent calculation for entanglement entropy \cite{2019arXiv191111977P}. The entropy resolvent for a bulk density matrix $\rho_{code}$ is defined as
\begin{equation}
	R_{\rho,ij}(\lambda) = \sum_{n=0}^{\infty} \frac{(\rho_{Rad}^n)_{ij}}{\lambda^{n+1}} ,
\end{equation}
where $\rho_R$ is the corresponding reduced density matrix on the radiation system $Rad$. Diagrammatically, we have
\begin{equation}
	\includegraphics[width = .9\textwidth,valign = c]{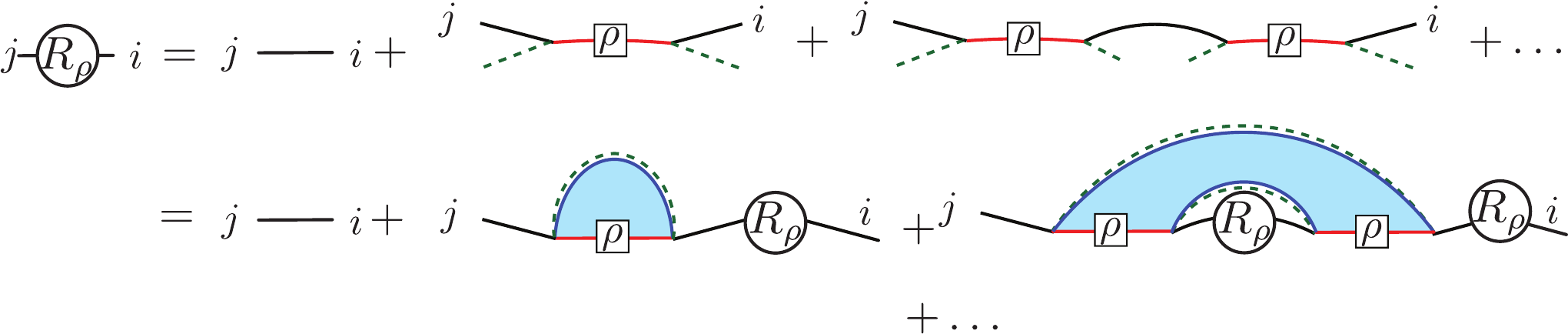},
\end{equation}
where, in the second line, the sum has been reorganized from an index running over the number of asymptotic boundaries to ab self-consistency equation indexed by the number of boundaries in the first connected component. 

This reorganization of the sum simplifies the resolvent and taking the trace of the above equation, one obtains \cite{2019arXiv191111977P,2021JHEP...04..062A}
\begin{equation}\label{eq:ent_res}
	\lambda R_{\rho}(\lambda) = k + \sum_{n=1}^{\infty} Z_n \tr(\rho_{code}^n)\frac{R_{\rho}(\lambda)^n}{k^n Z_1^n}.
\end{equation}
It is also useful to define a bulk entropy resolvent:
\begin{equation}
	R_{\rho}^{(\text{bulk})}(\lambda) = \frac{d_{code}}{\lambda}+ \sum_{n=1}^{\infty} \frac{\tr(\rho_{code}^n)}{\lambda^{n+1}}.
\end{equation}
The above expression allows us to write $\tr(\rho_{code}^{n})$ as a contour integral that picks out the relevant term in the series expansion of $R_{\rho}^{(\text{bulk})}(\lambda)$, i.e.,
\begin{equation}
	\tr(\rho_{code}^{n}) = \oint d\lambda_1\,\lambda_1^{n}R_{\rho:\sigma}^{(\text{bulk})}(\lambda_1),
\end{equation}
where the contour is chosen to wrap the pole at $|\lambda_1|=\infty$. 

Using these integral representations, one can sum the geometric series in \Eqref{eq:ent_res} to obtain
\begin{align}
	\lambda R_{\rho}(\lambda) &= k + \sum_{n=1}^{\infty} \left(\int_{0}^{\infty} ds\, \varrho(s)y(s)^n \right)\left(\frac{R_{\rho}(\lambda)^n}{k^n Z_1^n}\right) \left(\oint d\lambda_1\,\lambda_1^{n}R_{\rho}^{(\text{bulk})}(\lambda_1)\right)\\
	&=k + \int_{0}^{\infty} ds \oint d\lambda_1\,\frac{\varrho(s)R_{\rho}^{(\text{bulk})}(\lambda_1) R_{\rho}(\lambda) \lambda_1 w(s)}{k - R_{\rho}(\lambda) \lambda_1 w(s)}.
\end{align}
Solving this Schwinger-Dyson equation is generally hard, but suppose one can do so, the entropy resolvent can then be used to obtain the entanglement entropy. First, the density of eigenvalues is given by
\begin{equation}
	D(\lambda) = \lim_{\epsilon \rightarrow 0^+}\frac{1}{2\pi i}\left[R(\lambda-i\epsilon)-R(\lambda+i\epsilon)\right],
\end{equation}
and the entropy term is then given by
\begin{equation}
	\tr(\rho_{Rad} \log \rho_{Rad}) = \int_{0}^{\infty} D(\lambda) \lambda \log \lambda.
\end{equation}

One can define a similar resolvent for the relative term, which we call the relative resolvent:
\begin{equation}
	R_{\rho:\sigma,ij}(\lambda)= \frac{\delta_{ij}}{\lambda}+\sum_{n=1}^{\infty} \frac{(\rho_{Rad} \sigma_{Rad}^{n-1})_{ij}}{\lambda^{n+1}}.
\end{equation}
Using a similar diagrammatic representation, we have 
\begin{equation}
	\includegraphics[width = .9\textwidth,valign = c]{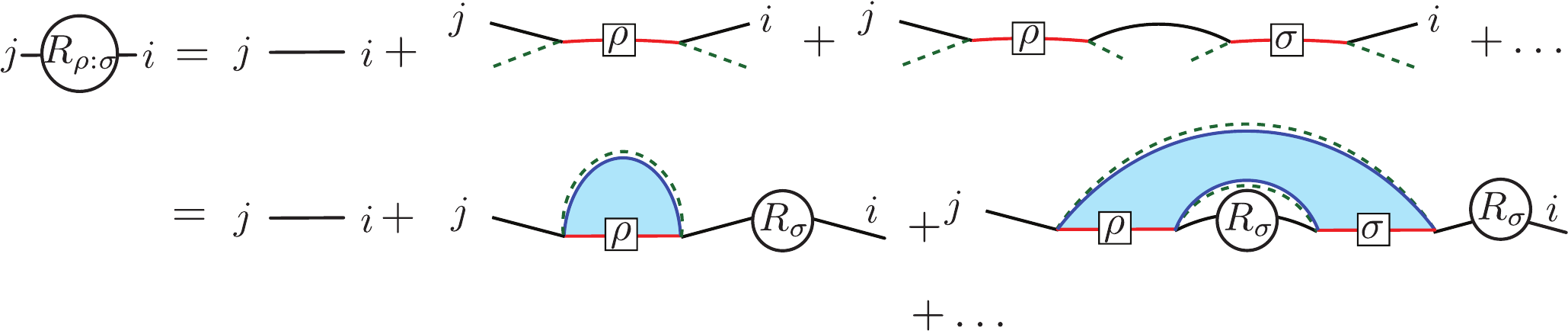},
\end{equation}
where we have again reorganized the sum to be indexed by the number of boundaries in the first connected component. Taking the trace of the above equation, one obtains
\begin{equation}\label{eq:rel_res}
	\lambda R_{\rho:\sigma}(\lambda) = k + \sum_{n=1}^{\infty} Z_n \tr(\rho_{code} \sigma_{code}^{n-1})\frac{R_{\sigma}(\lambda)^n}{k^n Z_1^n}.
\end{equation}

Given bulk density matrices $\rho_{code}$ and $\sigma_{code}$, one can define a bulk relative resolvent whose trace is given by
\begin{equation}
	R_{\rho:\sigma}^{(\text{bulk})}(\lambda) = \frac{d_{code}}{\lambda}+ \sum_{n=1}^{\infty} \frac{\tr(\rho_{code} \sigma_{code}^{n-1})}{\lambda^{n+1}}.
\end{equation}
The above expression allows us to write $\tr(\rho_{code} \sigma_{code}^{n-1})$ as a contour integral that picks out the relevant term in the series expansion of $R_{\rho:\sigma}^{(\text{bulk})}(\lambda)$, i.e.,
\begin{equation}
	\tr(\rho_{code} \sigma_{code}^{n-1}) = \oint d\lambda_1\,\lambda_1^{n}R_{\rho:\sigma}^{(\text{bulk})}(\lambda_1),
\end{equation}
where the contour is chosen to wrap the pole at $|\lambda_1|=\infty$. 

We can now use the above representation of $\tr(\rho_{code} \sigma_{code}^{n-1})$ and the integral representation of $Z_n$ in \Eqref{eq:rel_res} to obtain
\begin{align}
	\lambda R_{\rho:\sigma}(\lambda) &= k + \sum_{n=1}^{\infty} \left(\int_{0}^{\infty} ds\, \varrho(s)y(s)^n \right)\left(\frac{R_{\sigma}(\lambda)^n}{k^n Z_1^n}\right) \left(\oint d\lambda_1\,\lambda_1^{n}R_{\rho:\sigma}^{(\text{bulk})}(\lambda_1)\right)\\
	&=k + \int_{0}^{\infty} ds \oint d\lambda_1\,\frac{\varrho(s)R_{\rho:\sigma}^{(\text{bulk})}(\lambda_1) R_{\sigma}(\lambda) \lambda_1 w(s)}{k - R_{\sigma}(\lambda) \lambda_1 w(s)}.
\end{align}
Thus, we end up with a Schwinger-Dyson equation for the relative term. Although this equation is difficult to solve in general, it may be amenable to approximation techniques used in Ref.~\cite{2019arXiv191111977P}. Once we solve for $R_{\rho:\sigma}(\lambda)$, we have
\begin{equation}
	\tr(\rho_{Rad} \sigma_{Rad}^{n-1}) = \oint d\lambda R_{\rho:\sigma}(\lambda) \lambda^n,
\end{equation}
where we for integer $n$, we can deform the contour to wrap the real axis as described in Ref.~\cite{2019arXiv191111977P} for the Petz map. After doing this contour deformation, we can analytically continue $n$ to non-integer values and taking the $n$ derivative at $n=1$, we obtain
\begin{equation}
	\tr(\rho_{Rad} \log \sigma_{Rad}) = \int_{0}^{\infty} D_{\rho:\sigma}(\lambda) \lambda \log \lambda,
\end{equation}
where the \textit{relative distribution}, $D_{\rho:\sigma}(\lambda)$, is defined by
\begin{equation}
	D_{\rho:\sigma}(\lambda) =  \lim_{\epsilon \rightarrow 0^+}\frac{1}{2\pi i}\left[R_{\rho:\sigma}(\lambda-i\epsilon) - R_{\rho:\sigma}(\lambda+i\epsilon)\right],
\end{equation}
in analogy with the entropy resolvent. 

This pseudo-eigenvalue distribution may be of interest in its own right. Note that in the simple case where the bulk density matrices are pure states of fidelity $t$, \Eqref{relative_eig_distrib} implies that the relative distribution is related to the entanglement spectrum as
\begin{align}
    D_{\rho:\sigma}(\lambda) = \left(\frac{1-t}{k \lambda} + t\right) D(\lambda).
\end{align}

\section{General Tensors Networks: The Ford-Fulkerson Algorithm}
\label{app_TN}

Tensor networks have played a key role in understanding various key concepts in AdS/CFT related to quantum information. Tensor networks where the bulk tensors are drawn from random distributions \cite{2016JHEP...11..009H} have provided particularly fruitful results. We are thus motivated to solve for relative entropy in generality in random tensor networks. 

Consider a random tensor network with tensors labeled $i = 1, \dots, m$. The dimensions of the bonds connecting tensors $i$ and $j$ are denoted $d_{ij}$. There are also external bonds at the boundary of the network with dimensions $d_{i\partial}$. There may be bulk states in the tensor network, represented by additional bonds on the internal tensors. For the state $\rho$, we will always consider bulk states that can be formed from random unitaries. These can be pure states if the random unitary entangles the bonds from tensors in the bulk. They can also be mixed states if the unitary entangles the bonds with an auxiliary Hilbert space. For the state $\sigma$, the bulk state can be any state created by random unitaries on some bulk bonds and the maximally mixed state on the rest. When the bulk state is created by random unitaries, a random tensor is effectively added to the network.

For this class of tensor network states, we find
\begin{align}
    \Tr\left[\rho_R \sigma_R^{n-1} \right] = \frac{1}{\prod_{\{i,j \}}d_{ij}^n} \sum_{\{\tau_i \}} \prod_{\langle ij\rangle} d_{ij}^{C(\tau_i^{-1}\circ \tau_j)}\prod_{\langle i\partial_R\rangle} d_{i\partial_R}^{C(\eta^{-1}\circ \tau_i)}\prod_{\langle i\partial_{\bar{R}}\rangle} d_{i\partial_{\bar{R}}}^{C(\tau_i)}\prod_{\langle iMM\rangle} d_{iMM}^{C(\tau_i)}.
    \label{TN_srel_eq}
\end{align}
Here, $d_{iMM}$ represent the bond dimensions for the bulk degrees of freedom prepared in the mixed state for $\sigma$. $R$ represents a subsystem of the boundary and $\bar{R}$ its complement. There is a permutation $\tau_i$ for each tensor in the network. For the bulk tensors, this permutation runs over the entire $S_n$ permutation group, while for the tensors representing independent random unitaries preparing the bulk states of $\rho$ and $\sigma$, the sum is only over the $\mathbbm{1}\times S_{n-1}$ subgroup.

We now describe the derivation of the solution to \Eqref{TN_srel_eq} that uses the Ford-Fulkerson algorithm and provide an illustrative example. More details can be found in Refs.~\cite{2010JPhA...43A5303C, 2021PRXQ....2d0340K,2022JHEP...02..076K}. For permutation $\tau$, we define $|\tau|$ as the minimal number of transpositions needed to get from the identity element to $\tau$. This is related to the number of cycles by $|\tau| + C(\tau) = n$. Crucially, this satisfies a triangle inequality $|\tau_1 \tau_2| + |\tau_2^{-1}\tau_3| \geq |\tau_1 \tau_3| $. The condition for the triangle inequality to be saturated is for $\tau_1 \tau_2$ to be non-crossing in the cycles of $\tau_1 \tau_3$, which we denote $\tau_1 \tau_2 \leq \tau_1 \tau_3$. It is clear that maximizing the exponents in \Eqref{TN_srel_eq} is equivalent to minimizing
\begin{align}
    F_{\{\tau_i \}} = \sum_{\langle ij \rangle} w_{ij}|\tau_i^{-1} \circ \tau_j| +  \sum_{\langle i\partial_R \rangle} w_{i\partial_R }|\eta^{-1} \circ \tau_i| +  \sum_{\langle i\partial_{\bar{R}} \rangle} w_{i\partial_{\bar{R}}}|\tau_i| +  \sum_{\langle iMM \rangle} w_{iMM}|\tau_{i}|.
    \label{flow_min_eq}
\end{align}
Attaching a source to $\partial_{\bar{R}}$ and $MM$ and sink to $\partial_R$, we take augmenting paths through the network where the $w$'s are the capacities of the edges. Repeatedly applying the triangle inequality along the path $\tau_{i_k}$ for $k = 1\dots l$, where $l$ is the number of nodes along the path, we have
\begin{align}
    |\tau_{i_1}|+ |\tau_{i_1}^{-1}\circ \tau_{i_2}| + \dots + |\tau_{i_{l}}^{-1}\circ \eta|\geq |\eta| = n-1.
\end{align}
The left hand side of the equation is minimized, saturating the inequality, when the permutations are non-crossing and $\tau_{i_1} \leq \tau_{i_2} \leq \dots \leq\tau_{i_l}\leq \eta$. By repeated applying augmenting paths, we eventually arrive at a residual network where the source and sink are disconnected. This is an example of a maximal flow, $|f|$, and we have
\begin{align}
    F_{\{\tau_i \}} \geq (n-1)|f| + F_{\{\tau_i \}} (\mbox{residual network}).
\end{align}
It is then simple to minimize $F_{\{\tau_i \}} (\mbox{residual network})$ by examining for four terms in \Eqref{flow_min_eq} in the residual network. The first term asserts that any nodes that remain connected in the residual network that are not connect to the source or sink must have coinciding permutations. The second term asserts that all remaining nodes connected to the sink should be set to $\eta$. The final two terms assert that all remaining nodes connected to the source must be set to the identity. With these conditions, $F_{\{\tau_i \}} (\mbox{residual network}) = 0$, so we have solves for the conditions on $\{\tau_i\}$ to maximize the exponents of \Eqref{TN_srel_eq} and thus find the leading order relative entropy. These conditions were summarized in the main text. A completely analogous derivation can be made for the entropy term.

There is a subtlety if any of the nodes whose permutations only run over $\mathbbm{1}\times S_{n-1}$ are connected to the sink. In this case, we set the permutation to the next closest permutation, $\mathbbm{1}\times \eta_{n-1}$, which has $|\eta^{-1} \circ \mathbbm{1}\times \eta_{n-1}| = 1$. Thus, $F_{\{\tau_i \}}$ will additionally contain a term proportional to $(n-2)$. This term necessarily diverges when taking the replica limit
\begin{align}
    \lim_{n\rightarrow 1}\frac{1}{1-n} \log \left[ N^{2-n}\right] = \infty,
\end{align}
so the relative entropy diverges whenever this is the case.

We now explicitly work out an example involving a network with three tensors. As discussed in Ref.~\cite{2021JHEP...04..062A}, this can describe a state with two fixed-area surfaces with bulk fields present in each region. For demonstration, we consider the following bond dimensions
\begin{align}
\begin{tikzpicture}[scale = .9]
    \node[draw, shape=rectangle] (T3) at (0,0) {$T_3$};
    \node[draw, shape=rectangle] (T2) at (2,0) {$T_2$};
    \node[draw, shape=rectangle] (T1) at (4,0) {$T_1$};
    \draw [thick] (T3)  -- node[below] {$N^2$}(T2);
    \draw [thick] (T2) --  node[below] {$N^3$}(T1);
    \draw [thick, red] (T3) to node[left] {$N$} (0,1);
    \draw [thick, red] (T2) to node[left] {$N$} (2,1);
    \draw [thick, red] (T1) to node[left] {$N$} (4,1);
    \draw [thick] (T1) to node[below] {$N^4$} (6,0);
    \draw [thick] (T3) to node[below] {$N^4$} (-2,0);
    \end{tikzpicture},
\end{align}
where the red edges are bulk degrees of freedom, the right external edge is region $A$ and the left is region $B$. Considering the relative entropy between a random pure state in the code subspace, entangling the three bulk regions, and the maximally mixed state in the code subspace, we find the following flow network for the $\Tr\left[ \rho_R \sigma_R^{n-1}\right]$ term
\begin{align}
\begin{tikzpicture}[scale = .9]
    \node[draw, shape=rectangle] (T3) at (0,0) {$T_3$};
    \node[draw, shape=rectangle] (T2) at (3,0) {$T_2$};
    \node[draw, shape=rectangle] (T1) at (6,0) {$T_1$};
    \node[draw, shape=rectangle] (so) at (-3,0) {source};
    \node[draw, shape=rectangle] (si) at (9,0) {sink};
    \draw [thick] (T3)  -- node[above] {$2$}(T2);
    \draw [thick] (T2) --  node[above] {$3$}(T1);
    \draw [thick] (T3) --  node[above] {$4$}(so);
    \draw [thick] (T1) --  node[above] {$4$}(si);
    \draw [thick] (T3) to [in=-30,out=-150] node[below] {$1$} (so);
    \draw [thick] (T2) to [in=30,out=150] node[above] {$1$} (so);
    \draw [thick] (T1) to [in=-40,out=-140] node[below] {$1$} (so);
    \end{tikzpicture}.
\end{align}
The Ford-Fulkerson algorithm involves taking three augmenting paths: $\mbox{source}\rightarrow T_3\rightarrow T_2 \rightarrow T_1 \rightarrow \mbox{sink} $, $\mbox{source}\rightarrow T_2 \rightarrow T_1 \rightarrow \mbox{sink} $, and $\mbox{source} \rightarrow T_1 \rightarrow \mbox{sink} $. The total flow equals $4$. We are left with the following residual network
\begin{align}
\begin{tikzpicture}[scale = .9]
    \node[draw, shape=rectangle] (T3) at (0,0) {$T_3$};
    \node[draw, shape=rectangle] (T2) at (3,0) {$T_2$};
    \node[draw, shape=rectangle] (T1) at (6,0) {$T_1$};
    \node[draw, shape=rectangle] (so) at (-3,0) {source};
    \node[draw, shape=rectangle] (si) at (9,0) {sink};
    \draw [thick] (T3) --  node[above] {$3$}(so);
    \end{tikzpicture}.
\end{align}
The rules tell us that all permutations are non-crossing and $\mathbbm{1} = \tau_3\leq \tau_2 \leq \tau_1 \leq \eta$. The number of $\{ \tau_i\}$'s satisfying these conditions are known in the combinatorics literature to be given by the second Fuss-Catalan number \cite{2006math.....11106A}
\begin{align}
    FC^{(2)}_{n} :=  \frac{1}{2n +1}\binom{3n }{n}.
\end{align}
We thus find $\Tr\left[\rho_R \sigma_R^{n-1} \right] = FC_n^{(2)}N^{4(1-n)}$.

Moving on to the entropy term, we have a new flow network
\begin{align}
\begin{tikzpicture}[scale = .9]
    \node[draw, shape=rectangle] (T3) at (0,0) {$T_3$};
    \node[draw, shape=rectangle] (T2) at (3,0) {$T_2$};
    \node[draw, shape=rectangle] (T1) at (6,0) {$T_1$};
    \node[draw, shape=rectangle] (T4) at (3,-3) {$T_4$};
    \node[draw, shape=rectangle] (so) at (-3,0) {source};
    \node[draw, shape=rectangle] (si) at (9,0) {sink};
    \draw [thick] (T3)  -- node[above] {$2$}(T2);
    \draw [thick] (T2) --  node[above] {$3$}(T1);
    \draw [thick] (T3) --  node[above] {$4$}(so);
    \draw [thick] (T1) --  node[above] {$4$}(si);
    \draw [thick] (T3) to  node[below] {$1$} (T4);
    \draw [thick] (T2) to node[right] {$1$} (T4);
    \draw [thick] (T1) to  node[below] {$1$} (T4);
    \end{tikzpicture}.
\end{align}
This time, we need just two augmenting paths\footnote{While the choice of paths is not unique, the final answer following from the rules is.}: $\mbox{source}\rightarrow T_3\rightarrow T_2 \rightarrow T_1 \rightarrow \mbox{sink} $ and $\mbox{source} \rightarrow T_3\rightarrow T_4\rightarrow T_1 \rightarrow \mbox{sink} $. The total flow is $3$ and we are left with the following residual network
\begin{align}
\begin{tikzpicture}[scale = .9]
    \node[draw, shape=rectangle] (T3) at (0,0) {$T_3$};
    \node[draw, shape=rectangle] (T2) at (3,0) {$T_2$};
    \node[draw, shape=rectangle] (T1) at (6,0) {$T_1$};
    \node[draw, shape=rectangle] (T4) at (3,-3) {$T_4$};
    \node[draw, shape=rectangle] (so) at (-3,0) {source};
    \node[draw, shape=rectangle] (si) at (9,0) {sink};
    \draw [thick] (T2) --  node[above] {$1$}(T1);
    \draw [thick] (T3) --  node[above] {$1$}(so);
    \draw [thick] (T1) --  node[above] {$1$}(si);
    \draw [thick] (T2) to node[right] {$1$} (T4);
    \end{tikzpicture}.
\end{align}
The rules now assert that $\tau_3 = \mathbbm{1}$ and $\tau_1 = \tau_2 = \tau_4 = \eta$, so $\Tr\left[\rho_R^n \right] = N^{3(1-n)}$. All together, we have
\begin{align}
    D(\rho_R || \sigma_R) = \lim_{n\rightarrow 1}\frac{1}{1-n}\left(\log\left[FC^{(2)}N^{4(1-n)}\right]-\log\left[N^{3(1-n)}\right]\right) = \log\left[N \right]-\frac{5}{6}.
\end{align}
The correction to the JLMS formula here is the $O(1)$ term.

\addcontentsline{toc}{section}{References}

\providecommand{\href}[2]{#2}\begingroup\raggedright\endgroup


\end{document}